\documentclass[fleqn,usenatbib]{mnras}

\usepackage{newtxtext,newtxmath}

\usepackage[T1]{fontenc}
\usepackage{ae,aecompl}

\usepackage{graphicx}	
\usepackage{amsmath}	

\usepackage{amssymb}	
\usepackage{multicol}
\usepackage{bm}
\usepackage{booktabs}
\usepackage{cleveref}
\usepackage{natbib}
\usepackage{orcidlink}
\usepackage{adjustbox}

\newcommand{\secpoint}{\mbox{$''\mskip-7.6mu.\,$}}
\newcommand{\angstrom}{\mbox{\normalfont\AA}}

\title[{AURORA: Electron Density}]{The AURORA Survey: The Evolution of Multi-phase Electron Densities at High Redshift}

\author[M. W. Topping]{\orcidlink{0000-0001-8426-1141}Michael W. Topping\thanks{michaeltopping@arizona.edu}$^1$,
\orcidlink{0000-0003-4792-9119}Ryan L. Sanders$^2$,
\orcidlink{0000-0003-3509-4855}Alice E. Shapley$^{3}$,
\orcidlink{0000-0003-4464-4505}Anthony J. Pahl$^{\thanks{Carnegie Fellow}, 4}$,\newauthor
\orcidlink{0000-0001-9687-4973}Naveen A. Reddy$^5$,
Daniel P. Stark$^6$,
\orcidlink{0000-0002-4153-053X}Danielle A. Berg$^7$,
\orcidlink{0000-0003-1249-6392}Leonardo Clarke$^3$,\newauthor
\orcidlink{0000-0002-3736-476X}Fergus Cullen$^{8}$,
\orcidlink{0000-0002-1404-5950}James S. Dunlop$^{8}$,
\orcidlink{0000-0001-7782-7071}Richard S. Ellis$^{9}$,
\orcidlink{0000-0003-4264-3381}N. M. F\"orster Schreiber$^{10}$,\newauthor
\orcidlink{0000-0002-8096-2837}Garth D. Illingworth$^{11}$,
\orcidlink{0000-0001-5860-3419}Tucker Jones$^{12}$,
\orcidlink{0000-0002-7064-4309}Desika Narayanan$^{13, 14}$,
\orcidlink{0000-0002-5139-4359}Max Pettini$^{15}$,\newauthor
\orcidlink{0000-0001-7144-7182}Daniel Schaerer$^{16}$
\\
$^{1}$Department of Astronomy / Steward Observatory, University of Arizona, 933 N Cherry Ave, Tucson, AZ 85721\\
$^{2}$Department of Physics and Astronomy, University of Kentucky, 505 Rose Street, Lexington, KY 40506, USA \\
$^{3}$Department of Physics \& Astronomy, University of California, Los Angeles, 430 Portola Plaza, Los Angeles, CA 90095, USA\\
$^{4}$The Observatories of the Carnegie Institution for Science, 813 Santa Barbara Street, Pasadena, CA 91101, USA\\
$^5$Department of Physics \& Astronomy, University of California, Riverside, 900 University Avenue, Riverside, CA 92521, USA\\
$^6$Department of Astronomy, University of California, Berkeley, Berkeley, CA 94720, USA\\
$^7$Department of Astronomy, The University of Texas at Austin, 2515 Speedway, Stop C1400, Austin, TX 78712, USA\\
$^{8}${Institute for Astronomy, University of Edinburgh, Royal Observatory, Edinburgh, EH9 3HJ, UK}\\
$^{9}$Department of Physics \& Astronomy, University College London. Gower St., London WC1E 6BT, UK\\
$^{10}$Max-Planck-Institut für extraterrestrische Physik (MPE), Giessenbachstr.1, D-85748 Garching, Germany\\
$^{11}$Department of Astronomy and Astrophysics, UCO/Lick Observatory, University of California, Santa Cruz, CA 95064, USA\\
$^{12}$Department of Physics and Astronomy, University of California, Davis, 1 Shields Avenue, Davis, CA 95616, USA\\
$^{13}${Department of Astronomy, University of Florida, 211 Bryant Space Sciences Center, Gainesville, FL 32611 USA}\\
$^{14}${Cosmic Dawn Center at the Niels Bohr Institute, University of Copenhagen and DTU-Space, Technical University of Denmark}\\
$^{15}$Institute of Astronomy, Madingley Road, Cambridge CB3 OHA, UK\\
$^{16}$Department of Astronomy, University of Geneva, Chemin Pegasi 51, 1290 Versoix, Switzerland}
\begin{document}

\label{firstpage}
\pagerange{\pageref{firstpage}--\pageref{lastpage}}
\maketitle
\begin{abstract}
We present an analysis of deep {\it JWST}/NIRSpec spectra of star-forming galaxies at $z\simeq1.4-10$, observed as part of the AURORA survey. We infer median low-ionization electron densities of $268_{-49}^{+45}~\rm cm^{-3}$, $350_{-76}^{+140}~\rm cm^{-3}$, and $480_{-310}^{+390}~\rm cm^{-3}$ at redshifts z$=2.3$, $z=3.2$, and $z=5.3$, respectively, revealing an evolutionary trend following $(1+z)^{1.5\pm0.6}$.
We identify weak positive correlations between electron density and star formation rate (SFR) as well as SFR surface density, but no significant trends with stellar mass or specific SFR. Correlations with rest-optical emission line ratios show densities increasing with $\rm [NeIII]\lambda3869/[OII]\lambda3727$ and, potentially, $\rm [OIII]\lambda5007/[OII]\lambda3727$, although variations in dust attenuation complicate the latter. Additionally, electron density is more strongly correlated with distance from the local BPT sequence than can be explained by simple photoionization models.
We further derive electron densities from the CIII] doublet probing higher-ionization gas, and find a median value of $1.4_{-0.5}^{+0.7}\times10^4~\rm  cm^{-3}$, $\sim30$ times higher than densities inferred from [SII]. This comparison suggests a consistent HII region structure across cosmic time with dense, high-ionization interiors surrounded by less dense, low-ionization gas. We compare measurements of AURORA galaxies to predictions from the SPHINX galaxy formations,
highlighting the interplay between residual molecular cloud pressure in young galaxies and feedback from stellar winds and supernovae as galaxies mature.

\end{abstract}

\begin{keywords}
galaxies: evolution -- galaxies: ISM -- galaxies: high-redshift
\end{keywords}

%
%
%
%
\section{Introduction} 
\label{sec:intro}

Constructing a robust model of galaxy formation and evolution requires a self-consistent picture of stars, gas, and dust across cosmic time. These components are intricately interconnected, such that constraining any component provides valuable insights into the broader context of the galaxy. The electron density of the interstellar medium (ISM) serves as a key diagnostic of its current conditions, reflecting properties such as ionization state, temperature, and geometry, while also preserving a record of the processes that have shaped the galaxy both internally and externally. Molecular clouds that collapse to form stars provide an initial set of conditions of the ISM, which is continually modified by internal mechanisms, including stellar feedback, supernovae (SNe), and the dynamic balance between gas heating and cooling. External influences, such as gas accretion from the circumgalactic medium (CGM) or interactions with other galaxies, further drive this evolution. By investigating how electron densities vary over time and among a diverse set of galaxy properties, we can obtain critical insights into the processes governing galaxy evolution.

In order to extract accurate properties of many components of galaxies, we often utilize observational indicators that indirectly probe the underlying physical properties (e.g., stellar mass, abundances, attenuation, ionization conditions).
Due to the complex interplay between many of these properties \citep[e.g.,][]{Steidel2016}, investigating each aspect of galaxies often requires a host of underlying assumptions. 
One such quantity that must be determined is the electron density of the ISM, which has wide-reaching effects on the observed properties of galaxies.
Collisional excitation and de-excitation processes that depend on the electron density can impact line fluxes, which must be considered in atomic calculations of elemental abundances and gas-phase conditions \citep[e.g.,][]{Osterbrock1989}.
Furthermore, as electron densities increase, they can affect the shape of the nebular continuum through the suppression of the 2-photon process \citep[e.g.,][]{Cameron2023}.
In high-redshift galaxies, the nebular continuum can comprise $\simeq50\%$ of the total continuum emission in the rest-frame UV for the youngest galaxies \citep[e.g.,][]{Byler2017, Topping2022, Katz2024}, such that its variation can fundamentally alter the shape of the spectral energy distribution (SED), and thus recovered galaxy properties.

Extensive observational campaigns using ground-based NIR spectrographs on 8m-class telescopes have revealed much about the ISM of galaxies at high redshift ($z>1$).
These efforts have typically utilized one or more emission-line doublets whose flux ratio is dependent on the occurrence of collisional excitation and de-excitations, which is a function of the electron density \citep[e.g.,][]{Osterbrock2006}.
The [OII]$\lambda\lambda3727,3729$ and [SII]$\lambda\lambda6717,6731$ doublets have received the most use due to the ease of detection among star-forming galaxies, and presence in the optical spectra and the near-IR (rest-frame optical) spectra of galaxies at $z\simeq1-3$.
The first electron densities inferred for HII regions at high redshift indicated that star-forming regions at $z\gtrsim1$ are considerably more dense than their local counterparts \citep[e.g.,][]{Erb2006, Hainline2009, Lehnert2009, Bian2010, Shirazi2014}.
Building on these early results, statistical samples of star-forming galaxies at $z\simeq1-3$ were characterized, where typical low-ionization electron densities of $200-300~\rm cm^{-3}$ became well established \citep[e.g.,][]{Steidel2014, Sanders2016, Kaasinen2017, Kashino2017, Harshan2020, Davies2021, Berg2022}.

While most constraints on ISM densities have utilized either the [OII] or [SII] doublet, these emission lines do not provide a complete picture of the ISM.
Indeed, the [OII] and [SII] lines primarily trace low-ionization gas ($\simeq10$ eV) that may not represent that of entire HII regions.
Other emission lines exist in the rest-frame UV and optical that are sensitive to gas properties in regions of the HII regions existing at a higher ionization state. For example, the CIII]$\lambda\lambda1907,1909$ and SiIII]$\lambda\lambda1883,1892$ doublets require ionization energies of $\simeq16-24$ eV, while NIV]$\lambda\lambda1483,1486$ and [ArIV]$\lambda\lambda4711,4740$ are sensitive to yet higher energies ($>40$ eV).
As ionizing radiation fields become harder at high redshift, these higher-ionization probes may become increasingly important. 
By combining densities inferred for multiple phases of the ISM the structure and geometry of HII regions can be constrained.
Although density measurements inferred from low-ionization [SII] and [OII] lines far outnumber those from higher-ionization species, 
high-ionization lines (e.g., CIII]) indicate the presence of gas with significantly higher electron densities \citep[e.g.,][]{James2014, Berg2021, Mingozzi2022, Mainali2023}.

Prior to {\it JWST}, the redshift frontier of electron densities based on optical lines extended to $z\simeq 5$ beyond which [OII] is not accessible from the ground. While [OII] was detected in some systems at $z\simeq4$ \citep[e.g.,][]{Shapley2017, Witstok2021}, any resulting density constraints can lack crucial context from features at longer wavelengths (e.g., H$\alpha$, [OIII]$\lambda5007$, [OIII]$\lambda4363$, [NII]$\lambda6584$) that provide insight into stellar and gas-phase properties of galaxies.
Recently, {\it JWST} has significantly advanced our understanding of electron densities and other gas-phase properties at high redshift \citep[e.g.,][]{Tang2023,Fujimoto2023,Reddy2023b, Shapley2023, Shapley2024, Sanders2024, Isobe2023, RobertsBorsani2024, Nakajima2023, Abdurrouf2024, Yanagisawa2024, Hu2024}.
The electron densities of small numbers of galaxies up to $z\simeq10$ have been determined using rest-optical line ratios, and have indicated that the ISM of early galaxies may have an even higher density than at $z\simeq2$ on average \citep[e.g.,][]{Isobe2023, Reddy2023b, Abdurrouf2024, Li2024}. 
While these initial results are undoubtedly intriguing, they generally lack sample statistics or homogeneity in order to provide consensus on the average densities at very high redshift, or to connect densities with other galaxy properties. 
However, for the small number of existing constraints, simultaneous determinations of density along with other physical conditions indicate that the density evolution may be the result of a combination of metallicity and geometrical effects \citep[e.g.,][]{Abdurrouf2024} and gas densities that span much larger (kpc) scales \citep[e.g.,][]{Shirazi2014, Shimakawa2015, Bian2016, Jiang2019, Davies2021, Reddy2023a}.
Finally, {\it JWST} has improved prospects of detecting the weaker lines needed to constrain densities of the highly-ionized regions of the ISM. One consequence is the discovery of a small number of objects with exceptionally high densities \citep[$\gtrsim10^5~\rm cm^{-3}$; e.g.,][]{Bunker2023, Cameron2023, Senchyna2024, Topping2024a, Topping2024c}.
With {\it JWST}, it is now possible to disentangle the effects of electron densities and other gas-phase properties on the observed emission from galaxies, and to trace their impact across cosmic time.

In this paper, we investigate the electron densities inferred for galaxies observed during Cycle 1 of {\it JWST} operations as part of the Assembly of Ultra-deep Rest-optical Observations Revealing Astrophysics (AURORA) survey.
We utilize ultra-deep medium-resolution spectra with continuous wavelength coverage from $1-5\rm \mu m$ to constrain electron densities for a majority of the 97 galaxies targeted by the survey, which span a wide range in redshift and galaxy properties (Section~\ref{sec:properties}).
We investigate trends between the inferred densities with integrated galaxy properties such as star-formation rate (SFR) and stellar mass, and with emission-line ratios that are commonly used to extract properties from high-redshift galaxies. 
Finally, we compare the results from AURORA to predictions from cosmological radiation-hydrodynamic simulations to investigate the physical processes that drive these trends and the redshift evolution of electron densities.

The structure of this paper is as follows.
Section~\ref{sec:data} outlines the AURORA survey and describes the methodology to obtain final spectra as well as the physical properties inferred for the AURORA sample.
In Section~\ref{sec:results} we provide the main results of our analysis.
Finally, we provide a discussion of the results and brief summary in Sections \ref{sec:disc} and \ref{sec:summary}, respectively.
Throughout this paper we assume a cosmology with $\Omega_{\rm m} = 0.3$, $\Omega_{\Lambda}=0.7$, $H_0=70 \textrm{km s}^{-1}\ \textrm{Mpc}^{-1}$, and adopt solar abundances from \citet[][i.e., $Z_{\odot}=0.014$, $12+\log(\rm O/H)_{\odot}=8.69$]{Asplund2009}. All magnitudes are provided using the AB system \citep{Oke1984}.
%
%
%
%

\section{Data and Sample Properties}
\label{sec:data}

The spectroscopic sample used in this analysis was assembled as part of the AURORA survey (GO-1914; PI: Shapley).
The observational strategy and target selection criteria for AURORA are described in detail in \citet{Shapley2024} and \citet{Sanders2024}.
Briefly, the total AURORA sample comprises 97 high-redshift ($z>1.4$) galaxies using {\it JWST}/NIRSpec in two Multi-Shutter Assembly (MSA) configurations, one each in COSMOS and GOODS-N fields \citep{Grogin2011}.
The primary targets on the MSA ($N=36$) span a redshift range of $z=1.4-4.4$, and were identified based on detectable predicted fluxes of key auroral emission lines (e.g., [OIII]$\lambda4363$, [OII]$\lambda\lambda$7320,7330).
The remaining targets were assigned slits, in order of decreasing priority, based on the following categories: very high redshift ($z>6$) galaxies, $z>2$ quiescent galaxies, strong line emitters, and $z_{\rm phot}>1.5$ galaxies. These additional targets were primarily identified from the PRIMER survey (GO-1837; \citealt{Dunlop2021}, Dunlop et al. in prep), the FRESCO survey \citep{Oesch2023}, the JADES survey \citep{Hainline2024, Eisenstein2023}, and from the literature \citep{Jung2020, Finkelstein2015, Bouwens2015, Heintz2024}.
Each of the NIRSpec MSAs were observed with the G140M/F100LP, G235M/F170LP, and G395M/F290LP grating/filter combinations, with exposure times of 12.3, 8.0, and 4.2 hours, respectively, yielding continuous wavelength coverage from $1-5\rm \mu m$ for each target.

Section~\ref{sec:spectra} describes the instrumental setup and reduction of the spectroscopic data, in addition to supplementary imaging. Section~\ref{sec:resolution} discusses the possibility of resolving close doublets in the NIRSpec spectra, and Section \ref{sec:properties}  describes the methodology for inferring galaxy properties. Finally, Section \ref{sec:densities} defines the complete sample analyzed in this work.

\subsection{Data Reduction and Measurements}
\label{sec:spectra}
We followed the same reduction procedures to reduce the NIRSpec/MOS data obtained in all three gratings to obtain the final 2D spectra.
We began by passing the individual uncalibrated detector exposures through the {\it JWST} \texttt{calwebb\_detector1} pipeline\footnote{\url{https://jwst-pipeline.readthedocs.io/en/latest/index.html}}.
This step implemented a masking of all saturated pixels, subtraction of signal due to bias and dark current, and masking of `snowballs' and `showers' resulting from high-energy cosmic ray events.
The resulting images from this first reduction step were then corrected for striping by estimating and subtracting the $1/f$ noise in each image.
Following these calibrations, we cut out the 2D spectrum for each slit on the MSA and applied a flat-field correction, photometric calibration, and applied the wavelength solution using the up-to-date calibration reference data system (CRDS) context (\texttt{jwst\_1027.pmap}).
Each cutout slitlet was then rectified and interpolated onto a common wavelength grid for its grating and filter combination.
Finally, the calibrated cutout spectra were combined following the defined three-shutter dither pattern, while excluding pixels that had been masked in a previous step of the reduction.
The 2D error spectra were then calculated as the square-root of the combined variance from Poisson noise, read noise, flat fielding, and variance between exposures, summed in quadrature. 
Finally, we manually extracted both science and error spectra using an optimal profile \citep{Horne1986}.

We leveraged existing {\it JWST}/NIRCam and {\it Hubble Space Telescope (HST)} imaging in the COSMOS and GOODS-N fields to extract broadband photometric and morphological parameters for the AURORA sample.
Specifically, 93/97 of the targeted objects fall within the footprint of JADES in GOODS-N \citep{Eisenstein2023} and PRIMER in the COSMOS field (GO-1837, Dunlop et al. in prep).
We obtained photometry from the publicly available catalogs from the DAWN {\it JWST} Archive. These catalogs contain measurements in the F435W, F606W,
F814W, F850LP, F105W, F125W, F140W, F160W filters from HST/ACS and WFC3, and F090W, F115W, F150W, F200W, F277W, F356W, F410M, and F444W from {\it JWST}/NIRCam in the COSMOS field, while GOODS-N objects also have coverage in {\it HST} Advanced Camera for Surveys (ACS) F775W and {\it JWST}/NIRCam F182M, F210M, and F335M filters.
We measured galaxy sizes for each object following the methodology of \citet{Pahl2022}, that uses \texttt{galfit} \citep{Peng2002, Peng2010} to fit S\'ersic profiles to cutouts of each object. For objects fit by a single S\'ersic profile, we assigned a circularized effective radius, $r_e$, as $r_e\equiv r\sqrt{b/a}$, where $r$ is the half-light radius along the major axis and $b/a$ is the axis ratio. 
In cases where objects comprise multiple S\'ersic components, the effective radius was derived by first calculating the area covered by the brightest model pixels that comprise half of the total emission from the object, $A$, with the effective radius following from $A=\pi r_e^2$.

Finally, we applied several corrections to the extracted 1D spectra of each object.
We first calculated wavelength-dependent slit losses following the method outlined in \citet{Reddy2023b} to correct for emission not captured by the NIRSpec/MSA microshutters. The corrections were derived using galaxy morphologies that were convolved with a kernel to match the wavelength-dependent PSF of {\it JWST}/NIRSpec. A multiplicative correction factor was then calculated for each object based on the amount of emission that fell outside the extraction window of each spectrum, and both the 1D spectrum and the error spectrum were then corrected by this factor.
A penultimate flux calibration step was applied to the 1D spectra to address differences in the relative fluxes between gratings for the same objects.
This procedure is described in detail in \citet{Sanders2024}, however we provide an overview here. Briefly, we compared the flux normalization between adjacent gratings based on the continuum level and emission-line fluxes that lie at overlapping wavelengths.
On average, the relative flux normalizations we derived for G140M/G235M and G395M/G235M were 0.96 and 0.80, respectively, for the COSMOS observations and 0.96 and 1.02 for GOODS-N on average. 
Galaxies that were significantly more discrepant (i.e., outside the 1$\sigma$ scatter in relative flux normalization) were corrected based on their own inter-grating normalizations, while the rest of the sample was corrected with the average values presented above.
We then applied a final, absolute flux calibration to each spectrum that was determined by comparing the observed photometry with broadband fluxes calculated by passing the spectrum through the corresponding filter curves.

\begin{figure*}
    \centering
     \includegraphics[width=1.0\linewidth]{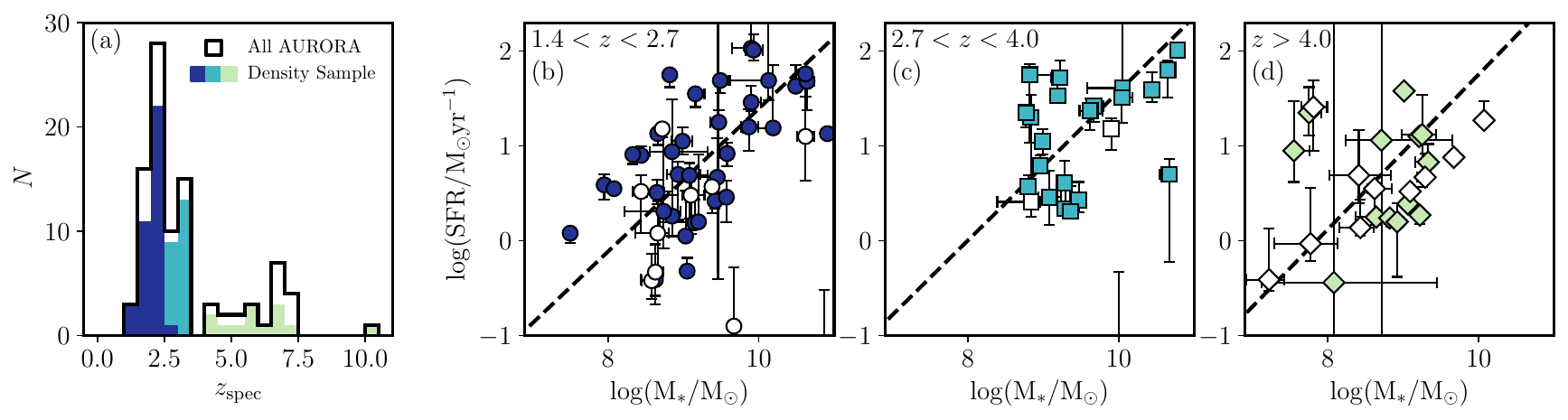}
     \caption{Sample statistics and galaxy demographics for objects analyzed in this work from the AURORA survey. The left panel presents the redshift distribution for the full AURORA survey (black histogram), as well as the $1.4<z<2.7$ (blue), $2.7<z<4$ (teal), and $z>4$ (green) electron density samples (see Section~\ref{sec:sample}). The right three panels present the SFR and stellar masses for the $1.4<z<2.7$ (left), $2.7<z<4$ (middle), and $z>4$ (right) electron density samples. In each SFR versus stellar mass diagram we overplot the star-forming main sequence parameterized by \citet{Speagle2014} at the median redshift of the sample. In each of the three redshift bins, the density samples scatter about the main sequence, indicating that they are representative of galaxies during their respective epochs.} 
     \label{fig:sample}
\end{figure*}

Emission lines were measured by fitting individual Gaussian profiles centered at the vacuum wavelength of each feature, and closely-spaced emission lines were fit with multiple profiles simultaneously. Furthermore, closely-spaced doublets of the same atomic species (e.g., [SII]$\lambda\lambda6717,6730$, [OII]$\lambda\lambda3727,3729$, CIII]$\lambda\lambda1907,1909$) were fit while requiring the line widths of both doublet members to be the same. Emission-line flux uncertainties were determined by perturbing the spectrum by its corresponding error spectrum and fitting Gaussian profiles to the result. The flux uncertainty was then set as the standard deviation of the distribution obtained repeating this perturbation process 1000 times.
In total, we measured emission lines, and thus obtained spectroscopic redshifts for, 95 galaxies in the AURORA sample out of the 97 targeted objects. Figure~\ref{fig:sample}a presents a redshift histogram of the galaxies targeted as part of this survey for which spectroscopic redshifts were determined. They span redshifts of $z=1.4-10.4$ with a median redshift of $z_{\rm med}=2.6$.

\subsection{Resolution of the NIRSpec Spectra}
\label{sec:resolution}
Unfortunately, several density-sensitive doublets are closely spaced in wavelength such that they border on being unresolved by the $R\simeq1000$ gratings on NIRSpec.
However, recent analyses have demonstrated that the effective resolution of {\it JWST}/NIRSpec may be higher for objects that are very compact \citep[see e.g.,][]{deGraaf2023}.
Furthermore, the wavelength dependence of the NIRSpec resolution means that the redshift of a galaxy may impact whether a given doublet can be resolved.
Thus, we quantified the spectral resolution for each galaxy in our sample based on its morphology to determine which doublets can reliably be used to infer electron densities.
The doublets that are most commonly detected from galaxies in our sample are [OII], [SII], and CIII]. The [SII] doublet members are separated by $\simeq13\angstrom{}$, in the rest frame, and are thus easily resolved for every galaxy in our sample (see Figure~\ref{fig:examples}).
However, both [OII] and CIII] are separated by only 2-3 \angstrom{} in the rest frame, nominally requiring spectral resolutions of $R\gtrsim1500-2500$ to separate.

To determine whether a given emission line doublet could be resolved, we developed a systematic approach to calculate the instrumental resolution for each emission line and each object.
We began by computing the doublet separation in units of NIRSpec resolution elements assuming the standard dispersion curves provided by STScI \footnote{Available for download at \url{https://jwst-docs.stsci.edu/jwst-near-infrared-spectrograph/nirspec-instrumentation/nirspec-dispersers-and-filters}}.
These resolution curves were calculated assuming the sizes of the targets are arbitrarily large, such that they uniformly fill the NIRSpec shutters. As such, they represented an effective lower limit on the resolving power of the instrument.
We derived an accurate spectral resolution for each object using \texttt{msafit} \citep{deGraaf2023}. For this calculation, we simulated the NIRspec observations based on the center and position angle of the NIRSpec shutter, in addition to the best-fit S\'ersic model derived from the methodology outlined above (see Section~\ref{sec:spectra}). The result of this modeling was a wavelength-dependent spectral resolution from which we extracted values for [OII] and CIII] based on their observed-frame wavelengths.
From this calculation, we found that the $R\simeq1000$ NIRSpec gratings were not sufficient to resolve [OII] for any galaxies in AURORA. However, CIII] is resolved in the spectra of 8 galaxies (57\% of CIII] detections in the AURORA sample). Thus, we include these 8 galaxies in the sample described in Section~\ref{sec:densities}. We present examples of resolved CIII] profiles in Figure~\ref{fig:examples} for two objects that were included in our sample as a result of this analysis.

\subsection{Physical Properties and Measurements}
\label{sec:properties}
\subsubsection{SED Models and Galaxy Properties}
Integrated galaxy properties were inferred using a variety of SED fitting procedures. These are described in detail in \citet{Shapley2024}, however we provide a brief outline of our methodology here.
Broadband magnitudes were obtained from a combination of previous HST imaging and {\it JWST}/NIRCam imaging. Photometric catalogs in the targeted fields were collected from the Dawn {\it JWST} Archive\footnote{\url{https://dawn-cph.github.io/dja/index.html}} (DJA, \citealt{Heintz2024}) based on imaging obtained from the PRIMER survey (Dunlop et al. in prep) in COSMOS, and from JADES \citep{Eisenstein2023} and FRESCO \citep{Oesch2023} in GOODS-N as described above.
We inferred galaxy properties from the best-fit SEDs to the broadband photometry. The photometry was first corrected for nebular emission, including the subtraction of line fluxes based on the spectroscopic measurements and the nebular continuum inferred from the Balmer lines.
We fit SEDs to these nebular-corrected broadband fluxes using \textsc{fast} \citep{Kriek2009}.
The stellar templates were constructed based on the Flexible Stellar Population Synthesis (FSPS) models of \citet{Conroy2009}, and assembled with a delayed-tau star-formation history (SFH) of the form $\rm SFR(t)\propto t\times e^{-t/\tau}$, where $t$ is the time since the onset of star formation and $\tau$ is the characteristic SFR timescale.
Each galaxy in the sample was fit with two separate models. The first model implemented a \citet{Calzetti2000} dust-attenuation law and a stellar metallicity of $1.4Z_{\odot}$, while the second model utilized a SMC dust law \citep{Gordon2003} in addition to a stellar metallicity of $0.27Z_{\odot}$. We selected the appropriate SED model for each galaxy based on the fit that provided a lower $\chi^2$ value.
Star-formation rates were calculated from the H$\alpha$ luminosity.
To calculate an SFR, we required a $>3\sigma$ detection in both H$\alpha$ and H$\beta$, and corrected the H$\alpha$ flux based on the observed Balmer decrement (assuming an intrinsic value of H$\alpha$/H$\beta=2.79$, corresponding to Case B recombination with $T_e=15,000~K$) and assuming the \citet{Cardelli1989} extinction curve. We then converted the corrected H$\alpha$ flux to luminosity based on the spectroscopic redshift of each galaxy. 
Finally, we derived an SFR from the H$\alpha$ luminosity by using a conversion factor of 
 $\rm SFR/M_\odot yr^{-1}=10^{-41.37} L_{H\alpha}/erg~s^{-1}$ for objects best fit by the $1.4Z_{\odot}$ SED model, and $\rm SFR/M_\odot yr^{-1}=10^{-41.59} L_{H\alpha}/erg~s^{-1}$ for objects best fit by the $0.27Z_{\odot}$ SED model. These conversion factors were calculated from BPASS models \citep{Stanway2018} with an upper stellar mass cutoff of $100~\rm M_{\odot}$ \citep{Reddy2022}.

\begin{figure}
    \centering
     \includegraphics[width=1.0\linewidth]{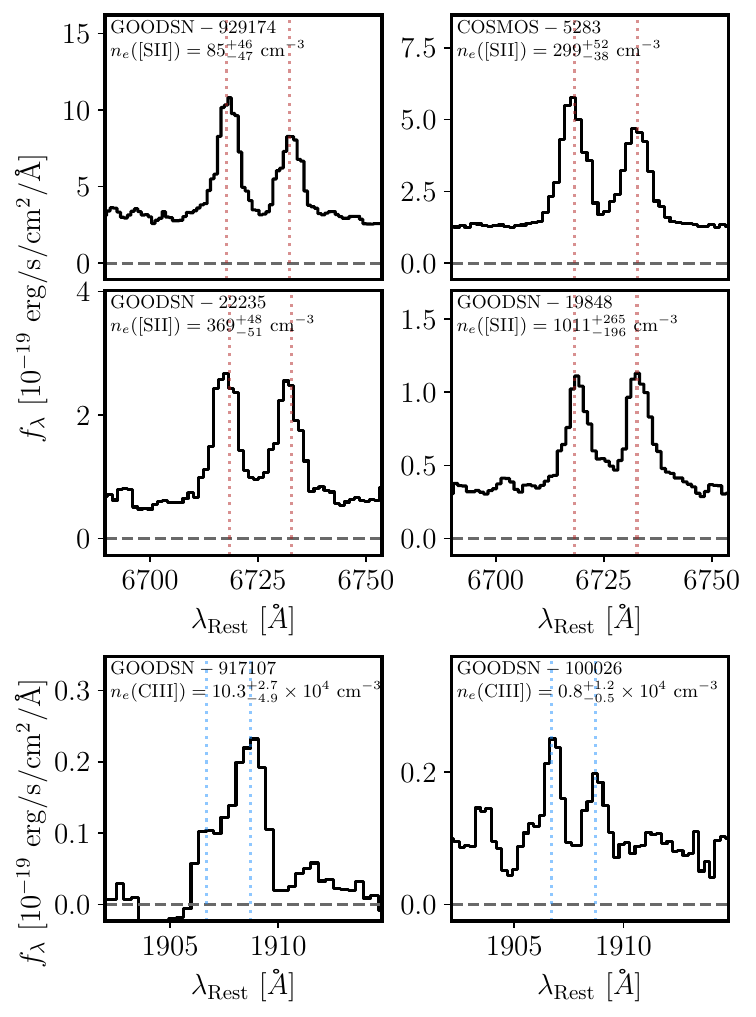}
     \caption{Example [SII] (top) and CIII] (bottom) line profiles for some objects in the AURORA sample. In each panel we list the object ID, as well as the electron density obtained following the methods described in Section~\ref{sec:densities}.} 
     \label{fig:examples}
\end{figure}

We compared the stellar masses inferred from the SED modeling described above to those obtained by fitting the broadband photometry using \textsc{prospector} \citep{Johnson2021} and assuming a non-parametric star-formation history (SFH).
First, we constructed stellar population SEDs using the FSPS models \citep{Conroy2009, Conroy2010} with a \citet{Chabrier2003} IMF and no nebular emission. We assembled models assuming a SFH comprising eight independent time bins constrained by a continuity prior \citep[see e.g.,][]{Tacchella2021}. The most recent bin of the SFH had a fixed width of 10 Myr, whereas the remaining bins were evenly distributed in log space from 10 Myr to the age of the Universe.
The priors for the properties of the dust content and metallicity were set to match those of the \textsc{fast} models. 
Models with SFHs that consider star-formation episodes over even shorter timescales (e.g., 3 Myr) may be able to more accurately reproduce the observed SEDs of galaxies currently experiencing a significant burst of star formation \citep[e.g.,][]{Whitler2023, Endsley2024}. Such flexibility is increasingly important at high redshift (i.e., $z\gtrsim6$), which describes only a small fraction of the AURORA sample. 
As an additional comparison, we utilized \textsc{prospector} to construct SED templates that self-consistently incorporate nebular continuum and line emission. We fit these templates to the photometry of our sample that had not been corrected for nebular emission. 
For this model setup, we used the same priors for the stellar population parameters described above, fixed the nebular and stellar metallicities to be identical, and allowed the ionization parameter to vary within $\log(U)\in[-4, -1]$ with a uniform prior.
The stellar masses inferred from these two \textsc{prospector} model setups are in good agreement with the results from the stellar-only \textsc{fast} models.
The \textsc{prospector} model fits that only considered stellar contributions yielded stellar masses that were 0.09 dex larger than from \textsc{fast}, with a scatter between masses from the two methods of 0.45 dex.
Similarly, fitting the broadband SEDs that include nebular emission with \textsc{prospector} provided stellar masses that were 0.15 dex larger than \textsc{fast}, with a scatter of 0.40 dex.

\begin{figure*}
    \centering
     \includegraphics[width=1.0\linewidth]{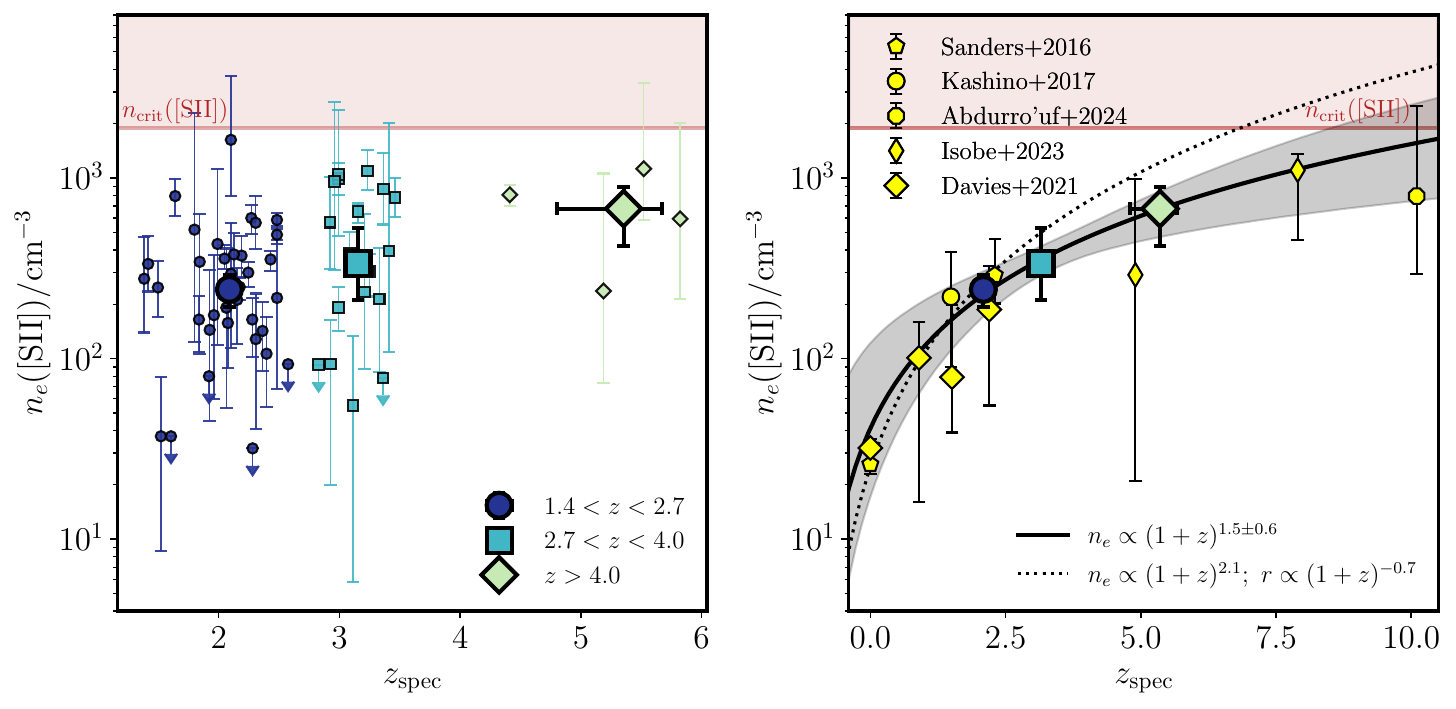}
     \caption{Left: Electron density inferred from the [SII] doublet (see Section~\ref{sec:data}) as a function of spectroscopic redshift for the aurora sample. The smaller points display values for individual objects, with the blue circles, teal squares, and green diamonds representing galaxies at $z<2.7$, $2.7<z<4$, and $z>4$, respectively. The larger outlined points indicate the median density within each redshift bin. Right: Comparison of the redshift evolution of electron densities within AURORA to typical values obtained from the literature based on either [SII] or [OII]. We include comparison points at or below the redshifts targeted by AURORA from \citet{Sanders2016, Kashino2017, Davies2021}, and include samples extending to higher redshifts from \citet{Isobe2023} and \citet{Abdurrouf2024}. The electron densities follow a monotonic trend of density increasing toward higher redshift, with the fastest evolution occurring between $z=0-2$, and a somewhat slower evolution up to $z\simeq10$. For reference, we display several power-law redshift evolution curves in black that are normalized to the density of the AURORA sample at $z=2$.} 
     \label{fig:density}
\end{figure*}

\subsubsection{Electron Density}
\label{sec:sample}
We infer electron densities from the observed doublet ratios using the python package PyNeb \citep{Luridiana2015}.
For all density calculations, we require that at least one doublet member is detected at $>5\sigma$. We place limits in cases where only one line is significantly detected.
When inferring densities based on doublet ratios, we assumed an electron temperature for the ISM of 15,000 K. This temperature is consistent with direct $T_e$ measurements determined at high redshift \citep[e.g.,][]{Sanders2024b, Kumari2024, Laseter2024}, and with other analyses that have constrained electron densities at high redshift \citep[e.g.,][]{Isobe2023, Abdurrouf2024}.
Electron temperatures in the range 10,000 K-20,000 K are commonly assumed, however adjusting the assumptions within this range has only a minimal effect on our results.
In addition to when there is a non-detection of one doublet member, we place limits on the inferred densities when the doublet ratio falls outside the range of values allowed by atomic physics. For [SII] this range is $0.45< \rm [SII]\lambda6716/\lambda6731 < 1.45$, while for CIII] the range of flux ratios is $0 < \rm CIII]\lambda1907/\lambda1909 < 1.5$. When a measurement falls beyond the high-density limit, we set the electron density to the critical density, and for emission line ratios beyond the low-density limit we set the electron density to the corresponding low-density limit of 10 $\rm cm^{-3}$ and 1000 $\rm cm^{-3}$ for [SII] and CIII], respectively.
Figure~\ref{fig:examples} presents [SII] and CIII] doublets for a small number of objects in the AURORA sample, and illustrates the variety of doublet ratios present for both lines.

\subsection{Electron Density Sample}
\label{sec:densities}
Armed with the set of derived quantities and measurements from the observed spectra, we constructed a sample of galaxies for which electron densities can be inferred.
Following the discussions in Sections~\ref{sec:resolution} and \ref{sec:sample}, we include galaxies in this sample that have a $>5\sigma$ detection in at least one member of a density-sensitive doublet, and that doublet must be resolved in the NIRSpec spectrum.
These requirements yield a sample of 51 galaxies with density constraints from [SII], and 8 galaxies with constraints from CIII]. Only 2 galaxies have density constraints from both lines, so the total number of galaxies with density constraints considered here is 57.

The redshift distributions of the full AURORA sample and the sample of objects for which we derive electron densities are presented in Figure~\ref{fig:sample}a. 
These objects span a redshift range of $z=1.4-10.4$, and roughly form a multi-modal distribution with peaks at $z\sim2$, $z\sim3$, and $z\sim5.5$. As such, we divide the sample into three redshift bins covering $z=1.4-2.7$, $z=2.7-4$ and $z>4$. This division of our sample follows the methodology presented in previous analyses as part of the AURORA survey \citep[e.g.,][]{Shapley2024}. 
The median redshift found in each bin is $z=2.1$, $z=3.2$, and $z=5.5$.

The right panels of Figure~\ref{fig:sample} present the stellar masses and SFRs for each AURORA redshift bin.
In each redshift range, the full AURORA sample scatters about the star-forming main sequence (SFMS) from \citet{Speagle2014}, implying that our sample is not systematically biased in SFR at fixed stellar mass, and is representative of typical galaxies at their respective epochs.
We further identified those galaxies for which we can derive electron densities and found that they are representative of the parent AURORA sample in each redshift bin.
We briefly summarize the galaxy properties of our electron density samples at $z\sim2.1$, $z\sim3.2$, and $z\sim5.5$ and derive median stellar masses (SFRs) of $\log(\rm M/M_{\odot})=9.1$ ($8.3~\rm M_{\odot}yr^{-1}$), $9.3$ ($19~\rm M_{\odot}yr^{-1}$), and $8.3$ ($7.0~\rm M_{\odot}yr^{-1}$), respectively.
While the AURORA galaxies are consistent with the SFMS at fixed redshift, we do find differences between the average properties of our samples within different redshift ranges.
Specifically, we find that galaxies at higher redshift are characterized by higher specific SFRs (sSFRs) following the evolution of the SFMS \citep[e.g.,][]{Speagle2014, Topping2022a, Popesso2023}.
We calculated average sSFRs of $3.7/\rm Gyr^{-1}$, $4.0/\rm Gyr^{-1}$, and $6.5/\rm Gyr^{-1}$ for the samples at $z\sim2.1$, $z\sim3.2$, and $z\sim5.5$, respectively.

\begin{figure*}
    \centering
     \includegraphics[width=1.0\linewidth]{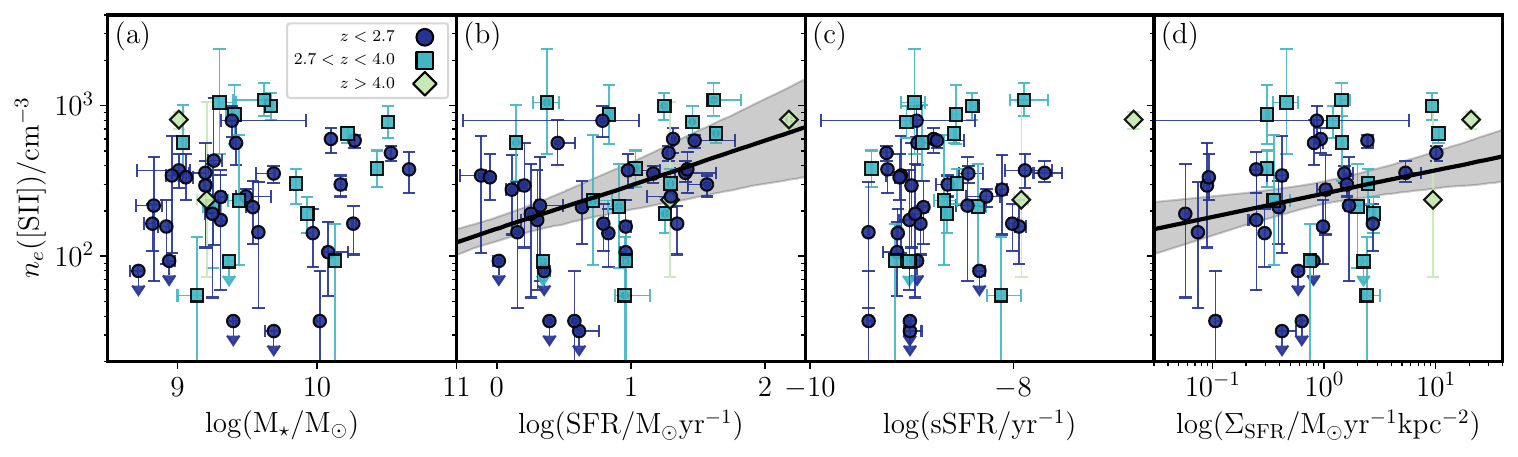}
     \caption{Electron density inferred from the [SII] doublet ratio displayed as a function of integrated galaxy SED properties. We compare the inferred densities to stellar mass (panel a), star-formation rate (panel b), specific-SFR (panel c), and SFR surface density (panel d). We display best-fit relations in the panels displaying SFR and $\Sigma_{\rm SFR}$ which are correlated with electron density at $\gtrsim2\sigma$ in the AURORA sample. These relations are presented in Equations 1 and 2, respectively.} 
     \label{fig:sedprops}
\end{figure*}

%
%
%
%
\section{Results}
\label{sec:results}

\subsection{Electron Densities of the low-ionization ISM from [SII]}

\subsubsection{Redshift Evolution}
\label{sec:evolution}

We begin by exploring how the low-ionization electron densities determined from [SII] vary among the AURORA sample.
Figure~\ref{fig:density}a presents the electron densities determined for each individual galaxy as well as in bins of redshift as described in Section~\ref{sec:densities}.
The lowest redshift sample ($z_{\rm med}=2.09^{+0.04}_{-0.09}$) in AURORA spans the largest range of electron densities out of the three bins, and contains galaxies with densities from the low-density limit (i.e., $\simeq10\rm ~cm^{-3}$) and up to the [SII] critical density. 
Among this sample of 37 $z\sim2$ galaxies we derive a bootstrapped median electron density of $268^{+45}_{-49}~\rm cm^{-3}$. This representative value is consistent with previous results of densities inferred from low-ionization transitions (e.g., [OII], [SII]) at $z\sim2$ \citep{Erb2006, Bian2010, Shirazi2014, Steidel2014, Sanders2016,Kaasinen2017,Davies2021, Reddy2023a}, which find typical electron densities in the range $200-300~\rm cm^{-3}$.

Galaxies in our sample at $2.7<z<4.0$ have a slightly higher electron density on average compared to those at $z\simeq2$.
For this redshift bin with a median redshift of $z_{\rm med}=3.2$, we derive a median electron density of $350^{+140}_{-76}~\rm cm^{-3}$. 
While there is considerable overlap in the [SII] densities between the systems at $z\simeq2$ and $z=3.2$, there are several differences between their density distributions.
First, the higher redshift bin contains a much higher fraction of systems with densities in excess of $650~\rm cm^{-3}$. Specifically, within the $z\simeq2$ bin only one such object is present, yet they comprise 35\% (7/20) of galaxies at $2.7<z<4.0$. 
At yet higher redshifts ($z>4.0$), the number of galaxies with electron densities that can be inferred from [SII] decreases significantly, and only four such objects are in our sample. The median redshifts of these [SII]-detected galaxies is $z_{\rm med}=5.3$.
This is due in part to the red cutoff wavelength of the NIRSpec G395M grating, such that the doublet is shifted out of the instrument coverage. Simultaneously, it is at these redshifts where density-sensitive doublets in the rest-UV become available to NIRSpec, which we discuss in detail in Section~\ref{sec:ciii}.
For the galaxies in AURORA with strong [SII] detections at $z>4$, we infer a median density of $480^{+390}_{-310}~\rm cm^{-3}$.
This density is higher than the medians found at both $z\simeq2$ and $z\simeq3.2$, however with a larger uncertainty owing to the small sample size. Notably, none of the galaxies at $z>4$ have low densities; each object in this redshift bin is consistent with or above the medians of the lower-redshift bins.

\begin{figure*}
    \centering
     \includegraphics[width=1.0\linewidth]{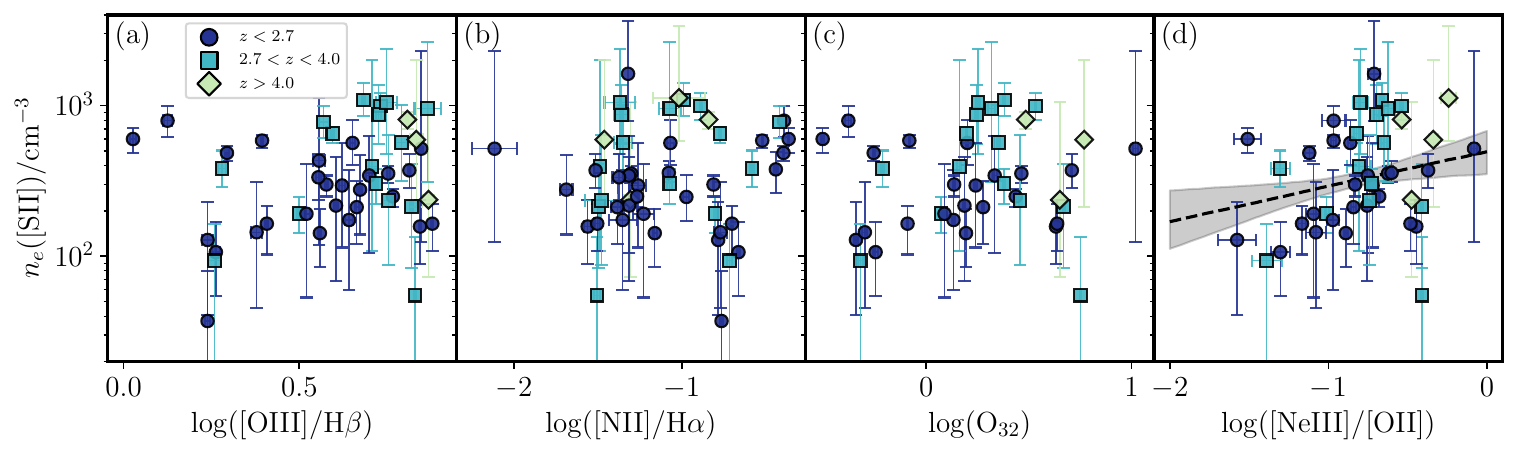}
     \caption{Electron density inferred from the [SII] doublet ratio versus rest-optical emission-line ratios for the AURORA sample. We compare the electron density to [OIII]/H$\beta$ (a), [NII]/H$\alpha$ (b), [OIII]/[OII] (c), and [NeIII]/[OII] (d). Emission line ratios that are sensitive to the ionization parameter (e.g., [OIII]/[OII], [NeIII]/[OII]) show a slight correlation with electron density, however several objects deviate from the average trend.} 
     \label{fig:lineprops}
\end{figure*}

With electron densities inferred for samples of galaxies at $z\simeq2.1$, $z\simeq3.2$, and $z\simeq5.5$, we can search for an evolutionary trend within the AURORA sample.
We fit a power-law to the median densities in these three bins and derive a best-fit evolutionary trend at a rate of $\propto(1+z)^{1.5\pm0.6}$. We present this best-fit trend along with the median densities from AURORA in addition to electron densities from the literature in Figure~\ref{fig:density}b.
We find that the best-fit trend extrapolated to higher redshift reaches an electron density of $n_{\rm e}\simeq900\rm ~cm^{-3}$ at redshifts of $z\ge 7.5$ which is comparable to the small numbers of existing measurements at that epoch \citep[e.g.,][]{Isobe2023, Abdurrouf2024}.
Conversely, the trend extrapolated toward lower redshift provides broad agreement with results at $z\le 2$; our best-fit trend yields an average density of $100^{+100}_{-60}\rm ~cm^{-3}$ at $z=1$, and a density of $40^{+60}_{-30}\rm ~cm^{-3}$ at $z=0$.
While estimates of the electron densities of star-forming galaxies at $z\simeq1-10$ scatter about our best-fit relation, the local values are systematically lower ($20-30\rm ~cm^{-3}$, \citealt{Sanders2016, Davies2021}), albeit consistent within the uncertainties.
This offset at low redshift may indicate a change in the evolution rate of typical electron densities. However, larger samples spanning a wider redshift baseline will be required to improve constraints throughout cosmic time.

\subsubsection{Integrated galaxy properties}

We next explore how the densities inferred from [SII] in the AURORA sample depend on integrated galaxy properties.
Figure~\ref{fig:sedprops} presents the [SII] densities as a function of stellar mass, SFR, specific SFR (sSFR), and star-formation rate surface density ($\Sigma_{\rm SFR}$).
We first consider the stellar masses and electron densities of the AURORA sample.
We find no significant monotonic trend between these two quantities among the full sample of $z\simeq1.4-5.8$ galaxies.
In addition, no correlation is present when the sample is restricted to galaxies within narrow redshift ranges of $z=1.4-2.7$ or $z=2.7-4$. The sample size of galaxies at $z>4$ with density estimates is not sufficient to constrain this correlation at higher redshifts. 
Nearly all galaxies in the AURORA sample have stellar masses within the relatively narrow dynamic range of $10^9-10^{10.5}~\rm M_{\odot}$; as a result, we may only be sensitive to very strong correlations between stellar mass and electron density. 
Finally, while we find no monotonic $\textrm{M}_\star-n_e$ trend, we note the lack of low densities ($<200/\rm cm^3$) at the high-mass end of the sample. 
We quantify this lack of low-density systems by dividing the AURORA sample into quartiles of stellar mass. The three lowest quartiles have median densities of 247, 225, and 223 $\rm cm^{-3}$, while the median density is roughly a factor of two higher for the most massive quartile (434 $\rm cm^{-3}$).

We next consider trends between electron density and quantities dependent on galaxy star-formation rates, beginning with SFR itself.
The AURORA galaxies do not show a significant ($>3\sigma$) correlation between SFR and electron density (Figure~\ref{fig:sedprops}b). However, we use the \texttt{linmix} package \citep{Kelly2007} to derive a tentative correlation at the $2\sigma$ level that is described by the functional form:
\begin{equation}
    \log(n_{\rm e}/\rm cm^{-3})=(0.29\pm 0.16)\times\log(\frac{\rm SFR}{M_{\odot}yr^{-1}})+(2.18\pm 0.13)~.
\end{equation}
This trend is primarily driven by the low-density boundary of the AURORA sample, which increases with increasing SFR (see Figure~\ref{fig:sedprops}b).
Indeed, all galaxies in our sample have densities that satisfy the relation: $\log( n_{\rm e}/\rm cm^{-3})>0.96\times \log(\rm SFR/M_{\odot}yr^{-1}) + 0.78$. 
In contrast, the highest densities reached by the sample do not depend on SFR; galaxies with densities consistent with $10^3~\rm cm^{-3}$ span the full range of SFR probed by AURORA galaxies.
In other words, this behavior comprises a narrowing of the electron density distribution as galaxies are restricted to higher densities as SFR increases.
Previous efforts to explore the correlation between SFR and density have yielded conflicting results. \citet{Sanders2016} similarly find no correlation between SFR and electron density \citep[see also][]{Shimakawa2015, HerreraCamus2016, Kashino2019, Davies2021} for a sample of 62 galaxies at $z\sim2.3$ from the MOSDEF survey that span a similar range of galaxy properties (e.g., stellar mass, SFR) to the AURORA sample. In contrast, \citet{Isobe2023} and  \citet{Shimakawa2015} find that the average density increases with increasing SFR.
A common result from larger samples is that densities are relatively insensitive to SFR at low SFR (0.1-10$\rm M_{\odot}/\rm yr$), and only increase above a sufficiently-high SFR threshold \citep[e.g.,][]{Davies2021}.
In this context, our findings are qualitatively consistent with previous results at $z\simeq2$, in that only at high SFRs does any pattern between SFR and $n_e$ emerge.
We find a similar behavior between density and sSFR.
While no significant monotonic relation between sSFR and density is present, the electron densities fill an envelope bounded by $\rm n_e < 1000 cm^{-3}$ and $\log(n_e)>0.65\times \log(\rm sSFR/yr^{-1}) + 7.31$.

Finally, we investigate how the electron densities depend on the star-formation surface density. 
We calculate the SFR surface densities ($\Sigma_{\rm SFR} = \frac{\rm SFR}{2\pi r^2}$) using the H$\alpha$ star-formation rates and effective radii as described in Section~\ref{sec:resolution} (see also \citealt{Pahl2022}).
The $\Sigma_{\rm SFR}$ and electron densities for the AURORA sample are presented in Figure~\ref{fig:sedprops}d. This quantity forms the most significant correlation with density out of those described in this section, however its significance is only at the $2.3\sigma$ level. We derive a best-fit linear relation of
\begin{equation}
        \log( n_{\rm e}\rm /cm^{-3})=(0.15\pm0.10)\times\log(\frac{\Sigma_{\rm \rm SFR}}{M_{\odot}yr^{-1}kpc^{-2}})+(2.42\pm0.08)
\end{equation}
that takes into account galaxies with density limits. In addition to this weak correlation, we find that galaxies in our sample with the highest (lowest) $\Sigma_{\rm SFR}$ also have a high (low) average electron density. In the AURORA sample, galaxies with $\Sigma_{\rm SFR}>5~\rm M_\odot/yr/kpc^2$ have an average density of $560^{+330}_{-210}~\rm cm^{-3}$, while those with $\Sigma_{\rm SFR}\leq 0.1~\rm M_\odot/yr/kpc^2$ have a considerably lower density of $200^{+130}_{-60}~\rm cm^{-3}$ on average.
The trend toward higher electron densities in galaxies with high SFR surface densities has been demonstrated in the local Universe \citep[e.g.,][]{Davies2021}, and at high redshift up to $z\simeq6$ \citep[e.g.,][]{Shimakawa2015, Sanders2016, Davies2021, Reddy2023a, Reddy2023b}.
The electron density of the ISM and the SFR surface densities are set by the properties of the parent molecular cloud. 
For the former quantity, an H II region is formed with the onset of star formation, where ionizing radiation photo-ionizes the molecular gas yielding an ISM with $n_e\simeq 2n_{\rm H_2}$, while the SFR surface density has been widely demonstrated to be related to molecular gas densities following the Kennicutt-Schmidt relation \citep[e.g.,][]{Hunt2009, Kashino2019}.
Thus, while it may not be surprising for these two quantities to be linked, feedback processes that persist over time can cause the electron density to stray from its initial value \citep[e.g.,][]{Oey1997}.

\begin{figure*}
    \centering
     \includegraphics[width=1.0\linewidth]{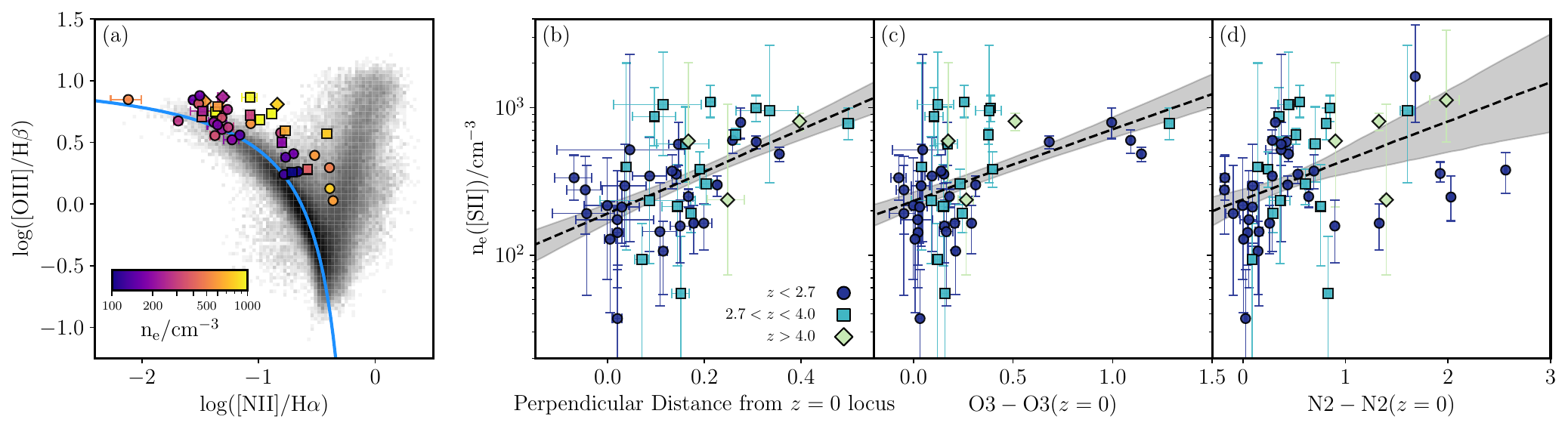}
     \caption{(a): Comparison between [SII] electron density and location on the BPT diagram. Galaxies in the AURORA sample are displayed as markers colored by their inferred densities, with circles, squares, and diamonds corresponding to galaxies at $z<2.7$, $2.7<z<4.0$, and $z>4.0$, respectively. The shaded black histogram illustrates local galaxies from SDSS, and the blue line presents a fit to the $z=0$ locus. (b): Electron density as a function of distance from the $z=0$ locus (in dex) for the AURORA sample. The marker shapes and colors distinguish the galaxies by their redshift bin. This comparison demonstrates that galaxies that are more offset from the local BPT sequence have higher electron densities on average. (c): Electron density as a function of offset in O3 line ratio at fixed N2 relative to the local sequence in dex. (d): Electron density as a function of offset in N2 from the local sequence at fixed O3 in dex.} 
     \label{fig:bpt}
\end{figure*}

\subsubsection{Empirical density trends with emission-line ratios}
\label{sec:lineprops}

Photoionization models used to reproduce the emergent spectra of high-redshift galaxies have shown a connection between the ratios of strong rest-optical emission lines and the electron density of the ISM \citep[e.g.,][]{Sanders2016, Steidel2014}.
\citet{Brinchmann2008} found a trend between emission line ratios and low-ionization electron densities among local SDSS galaxies when placed on the BPT diagram \citep{Baldwin1981}, such that galaxies with higher N2 ($\equiv \rm [NII]\lambda6584/H\alpha$) and O3 ($\equiv\rm [OIII]\lambda5007/H\beta$) ratios have higher electron densities.
This trend can be at least partially caused by increased collisional excitation driving higher N2 and O3 ratios \citep[e.g.,][]{Kewley2013}.
However, photoionization modeling of high-redshift HII regions suggest that the effect of density on emission line ratios and location on the BPT diagram location is less significant at lower metallicities where much of the high-redshift population resides.
In this section, we explore whether the change in emission line ratios with electron density is consistent with the density evolution derived in Section~\ref{sec:evolution}.

One complexity of comparing galaxy-integrated (derived from, e.g., photometry) and spectroscopic properties arises due to the comparatively small scales over which the NIRSpec spectra sample galaxies.
For the AURORA galaxies, the median circularized diameter is $0\secpoint3$ which is larger than the size of a NIRSpec shutter (0\secpoint2). Thus, the spectroscopic quantities (e.g., the electron density) do not fully capture the integrated properties of each galaxy for the majority of the AURORA sample.
The same complexity does not apply when considering the connections among different properties that have all been measured within the NIRSpec slit.
Specifically, we consider the ratios O3 ($\equiv \rm [OIII]\lambda5007/H\beta$), N2 ($\equiv \rm [NII]\lambda6583/H\alpha$), O32 ($\equiv \rm [OIII]\lambda5007/[OII]\lambda3727$), and Ne3O2 ($\equiv \rm [NeIII]\lambda3869/[OII]\lambda3727$), which trace a variety of properties including the metal content, abundance patterns, and ionization state. Critically, since both these properties and the electron densities are extracted from the same spectrum, the quantities are not affected by emission from outside the small region captured within the NIRSpec shutter that can impact photometric quantities.

We present the electron densities as a function of emission-line ratios in Figure~\ref{fig:lineprops}.
Of the emission-line ratios considered here, we find that only Ne3O2 presents a significant correlation with electron density, from which we derive the best-fit relation of:
\begin{equation}
    \log(n_e~\rm /cm^{-3}) = (0.24\pm0.10)\times\log(\rm Ne3O2) + (2.70\pm0.15)
\end{equation}
The Ne3O2 line ratio has been demonstrated to be closely related to the ionization parameter, U, \citep[e.g.,][]{Steidel2014, Steidel2016, Sanders2016a}, which in turn can be parameterized in terms of the electron density as:
\begin{equation}
    U\propto n_e^{1/3} \epsilon^{2/3},
\end{equation}
for an ionization-bounded nebula, where $\epsilon$ is the volume filling factor \citep{Brinchmann2008}.
Thus, we expect that any emission-line ratio that varies strongly with ionization parameter will also be sensitive to the density.
Such trends between ionization parameter and electron density have been established among samples of galaxies at $z\simeq1-6$ \citep[e.g.,][]{Reddy2023a}, where increased densities in galaxies with more extreme ionization conditions are seen.

Despite the trend between density and Ne3O2, we find that the AURORA sample lacks a significant correlation between density and O32, despite the fact that Ne3O2 and O32 themselves are strongly related \citep[e.g.,][]{Steidel2016, Tang2019}.
One key difference between these two line ratios is the impact of dust attenuation. While the close wavelength spacing of [NeIII] and [OII] makes Ne3O2 nearly immune to the effects of dust, an accurate determination of O32 requires a dust correction.
For the AURORA sample, the average correction we applied to the observed O32 ratios is 30\%, with the largest correction being a factor of $3$.
Any uncertainty of either the galaxy-to-galaxy dust curve or the Balmer line fluxes can propagate into increased scatter of the corrected O32 line ratios, which may be sufficient to disrupt the somewhat weak ($\rm O32\sim U \propto n_e^{1/3}$) correlation that is expected. 
In order to address these uncertainties, individualized dust corrections derived for each galaxy may be required (Reddy et al. in prep).

While neither N2 or O3 vary significantly with electron density individually, a trend arises when the two line ratios are considered in concert.
Specifically, we find that electron density is correlated with the position of galaxies on the $\rm [OIII]/H\beta$ vs. $\rm [NII]/H\alpha$ BPT diagram \citep{Baldwin1981}.
We quantify the BPT location of the AURORA sample in three different ways: N2 offset from the local sequence at fixed O3, O3 offset from the local sequence at fixed N2, and perpendicular distance from the local sequence. 
Figure~\ref{fig:bpt} compares the electron densities of the AURORA sample to these three parametrizations of offset from local galaxies on the BPT diagram.
In each case, electron density is significantly correlated with the BPT location. We quantify each trend as: 
\begin{equation}
\begin{cases}
    \log(n_{\rm e}/\rm cm^{-3}) = (0.20\pm0.07)\times(\rm N2-N2_{z=0}) + (2.4\pm0.1)\\
    \log(n_{\rm e}/\rm cm^{-3}) = (0.48\pm0.11)\times(\rm O3-O3_{z=0}) + (2.3\pm0.1)\\
    \log(n_{\rm e}/\rm cm^{-3}) = (1.44\pm0.28)\times(\rm \perp BPT_{z=0}) + (2.3\pm0.1)\\
\end{cases}
\end{equation}
where $\rm N2_{z=0}$ ($\rm O3_{z=0}$) is the value of [NII]/H$\alpha$ ([OIII]/H$\beta$) defined locally by \citet{Kewley2013}, and $\rm \perp BPT_{z=0}$ is the perpendicular distance from the local locus in dex.
While these best-fit relations are derived for the full AURORA sample, we note that similar trends are present among the individual redshift bins of our sample.
These correlations are qualitatively similar to the trends between emission-line ratios and density found from photoionization models described above.
However, low-metallicity models that change in density by an order-of-magnitude show only minor ($<0.1$ dex) changes in their rest-optical line ratios.
These small changes do not account for the range of densities seen in our sample (see Figure~\ref{fig:bpt}b), suggesting that other effects (e.g., hardness of the ionizing spectrum, abundances, ionization conditions) impact the line ratios of high-redshift galaxies.

\subsection{Densities from high-ionization lines}

\label{sec:ciii}
The HII regions within galaxies are thought to be composed of multiple phases of gas in differing ionization states \citep[e.g.,][]{Berg2021}.
In a simplified description of their structure, the most highly-ionized gas is situated toward the center of the HII regions near to the ionizing sources, while gas at increasing distances will comprise gas in a decreasingly ionized state.
Thus far we have considered electron densities inferred from [SII], which traces this low-ionization gas that exists on the outskirts of HII regions.
To probe the highly-ionized innermost gas, we require additional density-sensitive emission-line doublets, such as SiIII], CIII], or [ArIV].
These doublets require ionization energies of 16.3, 24.4, and 40.7 eV, in contrast to the 10.4 eV for [SII], and are at wavelengths that are available to {\it JWST}/NIRSpec at high redshift. 
Of these possibilities, CIII] is typically the strongest line seen in emission and is thus the most useful probe of density of the highly-ionization ISM.
At the rest-frame wavelength of CIII], we are only able to constrain it in the AURORA sample at $z>4.1$.
Consequently, a majority of the AURORA sample has no constraints on the CIII] emission. Nevertheless, we achieve $>5\sigma$ detections in both doublet members for a sample of 8 galaxies with redshifts of $z=5-11$ ($z_{\rm med}=6.3$; see also Section~\ref{sec:densities}).

\begin{figure}
    \centering
     \includegraphics[width=1.0\linewidth]{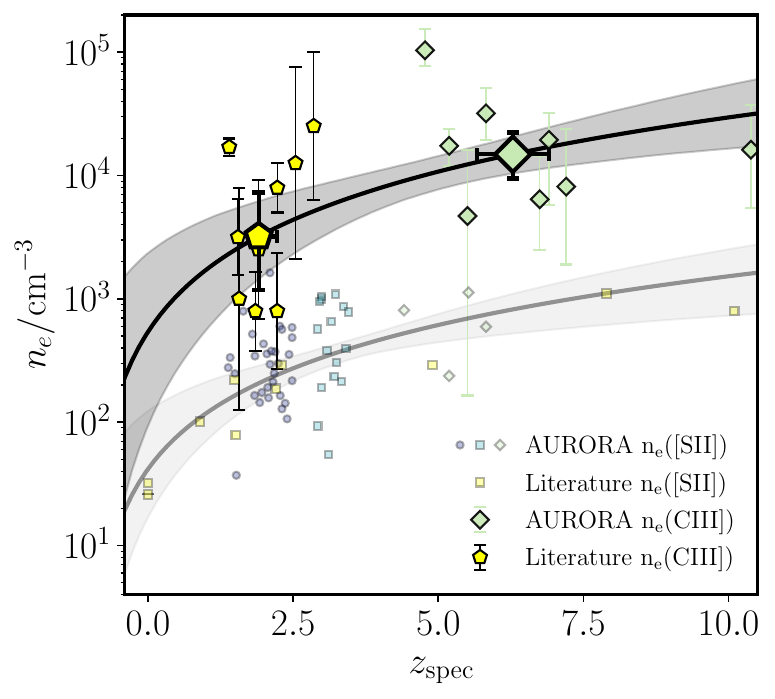}
     \caption{Redshift evolution of electron densities inferred from the CIII] doublet. Densities obtained for AURORA survey galaxies are displayed as green diamonds, while lower-redshift comparison galaxies from \citet{Maseda2017} are shown as yellow pentagons. In both samples, the median is presented as a larger symbol. We present densities inferred from [SII] (see Figure~\ref{fig:density}) from AURORA  as small transparent symbols. The black line and shaded region represents the best-fit evolutionary trend and corresponding uncertainty of $\propto(1+z)^{1.6^{+0.8}_{-0.7}}$.} 
     \label{fig:ciii}
\end{figure}

Figure~\ref{fig:ciii} presents the densities inferred from the CIII] doublet for AURORA galaxies where the line is significantly detected and resolved. 
All of the CIII] densities are considerably higher than those inferred from [SII] (Section~\ref{sec:evolution}).
We infer CIII] densities that range from $4\times10^3-10^5~\rm cm^{-3}$, with a median value of $1.4^{+0.7}_{-0.5}\times10^4~\rm cm^{-3}$. In comparison, the median density inferred from [SII] over the same wavelength range ($z>4$) is $480~\rm cm^{-3}$, representing a factor of 31 lower.
Due to the small overlap in the redshift range where both CIII] and [SII] are observable with NIRSpec, only two objects in our sample have densities inferred from both ions simultaneously. For these two objects, the CIII] densities exceed those of [SII] by a factor of $27$ on average, which is consistent with the comparison of the median densities at $z>4$. We return to explore the evolutionary trend of the CIII] densities in Section~\ref{sec:structure}.

Finally, we note that one object has a CIII] density that is a significant outlier in the AURORA sample.
We infer a density for GOODSN-917107 of $10^5~\rm cm^{-3}$, which is a factor of $6$ higher than the median CIII] density, and $3.5$ times greater than the second-highest CIII] density in the sample.
The [SII] doublet is covered, but not detected.
A small number of galaxies with similarly high densities in the ionized ISM have been identified with {\it JWST}, the first of which being GN-z11 \citep{Bunker2023, Senchyna2024} at $z=10.6$ followed by RXCJ2248-ID and A1703-zd6 at $z\simeq6-7$ \citep{Topping2024a, Topping2024c}.
Some defining features of these systems are a significant nitrogen enhancement, and the presence of strong high-ionization lines. While these features cannot be observed with {\it JWST}/NIRSpec at the redshift of GOODSN-917107 ($z=4.7$), this galaxy is similar to the $z>6$ objects in many respects.
In particular, GOODSN-917107 has large rest-UV line EWs ($\rm EW_{\rm CIII]}=44\pm14\angstrom{}$), a young age ($\rm \lesssim10 ~Myr$) inferred from the best-fit FAST SED, and is very compact (unresolved with NIRCam). This object may be a lower-redshift example of these dense, reionization-era systems, however additional measurements in the rest-frame UV will be required for confirmation.

\begin{figure*}
    \centering
     \includegraphics[width=0.95\linewidth]{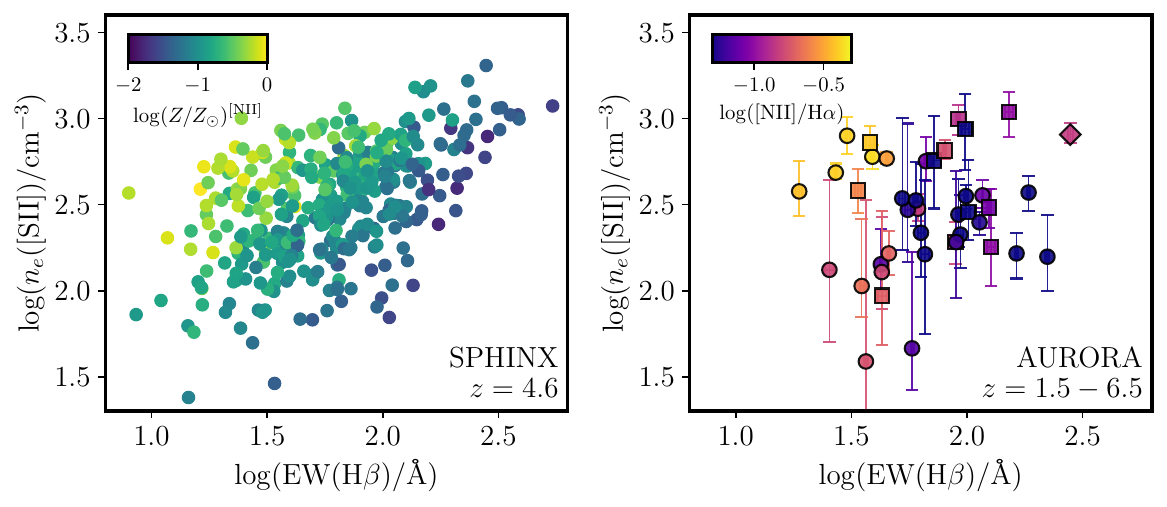}
     \caption{Comparison of electron densities and H$\beta$ EWs, where the latter is a proxy for recent star-formation activity. Left: Results from the SPHINX simulations at a $z=4.64$ snapshot \citep{Katz2023}. Simulated galaxies are color-coded based on their [NII]-weighted gas-phase metallicity. Right: Densities inferred from [SII] as a function of H$\beta$ EW from the AURORA survey, where each galaxy is color-coded by N2 ratio.} 
     \label{fig:hbew}
\end{figure*}

%
%
%
%
\section{Discussion}
\label{sec:disc}

\subsection{Density structure of high-redshift HII regions}
\label{sec:structure}
The co-evolution of the [SII] and CIII] densities that each probe different gas at different radii informs how the structure of HII regions evolves with cosmic time.
Due to the wavelength range of the {\it JWST}/NIRSpec spectra, AURORA can only constrain CIII] densities at $z\gtrsim4$, which limits our ability to explore changes with redshift.
To address this limitation, we consider additional star-forming galaxies at $z\sim2-3$ detected in CIII] from \citet{Maseda2017}. We compare these densities to the AURORA sample in Figure~\ref{fig:ciii}.
Just as in our high-redshift ($z>4$) comparison, the CIII] densities at $z\sim2$ are considerably higher than those inferred from [SII] \citep[see also,][]{James2014, Mainali2023}.
We find a median CIII] density from this comparison sample of $3300~\rm cm^{-3}$, which is a factor of $13$ higher than the AURORA [SII] measurements at the same redshift.

The CIII] densities at $z\sim2$ and $z\simeq6$ represent an evolutionary trend at a rate of $\propto(1+z)^{1.6^{+0.8}_{-0.7}}$, which is consistent with the evolution of [SII] densities.
We expect that the interiors of the HII regions are subject to more intense ionization conditions, such that CIII] is probing smaller radii in comparison to [SII] \citep[e.g.,][]{Berg2021}.
The similar rates of evolution of the [SII] and CIII] densities suggest that the structure of HII regions is similar at $z\sim2$ and $z\sim6$, but with an overall increase in density throughout the ISM.
If the electron densities are primarily driven by the geometry of galaxies, we may expect the overall mass density and the electron density to evolve at a similar rate.
\citet{Martorano2024} find that the radii of star-forming galaxies evolve at a rate of $\propto(1+z)^{-0.7}$, which corresponds to densities that evolve as $\rho \sim M/r^3\propto (1+z)^{2.1}$ at fixed stellar mass.
This rate represents a faster evolution than what we have derived for the electron densities inferred from both [SII] and CIII].
While we do not have the necessary sample to explore electron densities at fixed stellar mass from $z=0-10$, the AURORA samples at $z\simeq2.1$ and $z\simeq3.2$ are well-matched in stellar mass (see Section~\ref{sec:sample}).
Between these two redshifts, the median electron density in AURORA galaxies changes by $30\%$, while an evolution of $(1+z)^{2.1}$ corresponds to a change of nearly a factor of two over the same time period, suggesting that processes other than just the increased compactness of galaxies toward earlier times must be at play.
While this density gradient may be appropriate for HII regions that are ionization bounded, the [SII] and CIII] densities described above could also be represented by a density-bounded nebula where escaping ionizing radiation excites gas farther from the ionizing sources. 
However, the majority of the CIII] emission likely originates from dense clumps of gas, whereas the [SII] emission is produced by a more diffuse, low-density medium. 
Nevertheless, in both the density- and ionization-bounded scenarios, the gas that is responsible for the CIII] and [SII] emission are distinct components.

The comparisons between densities inferred from [SII] and CIII] demonstrate that the high-redshift ISM is not described by a single uniform component, and highlight the challenge in modeling and interpreting the spectra of high-redshift galaxies.
Specifically, simple constant-density models, which are commonly used in this context do not take into account the order-of-magnitude difference in density that we have observed in our sample.
The existence of different densities within the ISM also illustrates the need for multiple density indicators that probe the different regions of the ISM.
This is because there is little-to-no overlap in density where both the [SII] and CIII] line ratios are simultaneously not in either a low or high-density limit.
Therefore, the CIII] ([SII]) cannot be used to infer the electron density of gas that is responsible for the [SII] (CIII]) emission.
Because these two emission lines are sensitive to the conditions of different parts of the ISM, care must be taken to use the appropriate conditions when interpreting the observed emission from galaxies.

\subsection{Comparison with Simulations}

The previous section demonstrated that the HII regions within high-redshift galaxies are composed of gas at a wide range of densities. Because emission lines from these systems react differently to the distribution of gas densities, simplistic photoionization models may not accurately describe the emission of high-redshift galaxies.
Here, we compare galaxies in our sample with simulated galaxies from the SPHINX cosmological radiation-hydrodynamic simulations \citep{Katz2023}.
These simulations are designed to have the resolution required to capture the multiphase ISM and self-consistently calculate the emission from simulated galaxies, thus representing a much more realistic comparison for the AURORA galaxies.

Here, we focus on the H$\beta$ equivalent width, a key diagnostic that encodes information about both the age of a galaxy and the hardness of its ionizing spectrum, providing a comprehensive view of the stellar population. 
These properties influence the ionization structure and drive processes such as feedback, which regulate the density of the interstellar medium. Compared to properties derived from broadband SEDs, H$\beta$ EWs offer distinct advantages as indicators of stellar population properties. 
While SED fitting captures integrated galaxy-wide properties, the density inferred from NIRSpec spectra is localized to the region within the NIRSpec shutter. 
Since the shutters were preferentially placed on the brightest regions of each galaxy, the spectral properties may systematically differ from the integrated properties derived from SED fitting. 
Additionally, if galaxies host multiple stellar populations, the spectrum may fail to accurately represent an average galaxy age, further contributing to potential biases.

We display the electron density of the low-ionization gas as a function of H$\beta$ EWs for galaxies in the SPHINX cosmological radiation-hydrodynamic simulations in Figure~\ref{fig:hbew}a.
For this analysis, we consider the SPHINX at a snapshot of $z=4.64$, which is the lowest available redshift, and is within the range of redshifts spanned by the AURORA electron density sample.
The densities and H$\beta$ EWs of the simulated galaxies are highly correlated, with a probability of being uncorrelated of $p<10^{-20}$. We derive a best-fit relation between these two quantities of $\log(n_e) = 0.5\times \log(\rm EW/\angstrom{})+1.6$.
Alongside this positive correlation, we find that the distribution of densities narrows as the H$\beta$ EW increases. For low H$\beta$ EWs ($<30\angstrom{}$), the width of the electron density is $\sigma_{\rm n_e}=0.35$ dex, which decreases to just $0.12$ dex for the subsample of galaxies with the highest H$\beta$ EWs ($>200\angstrom{}$).
Instead of this change in distribution width being the result of an increased intrinsic scatter at low EW, it is caused by a secondary dependence on metallicity.
This secondary dependence shows that at fixed H$\beta$ EW, higher metallicity galaxies have higher electron densities, or conversely, galaxies with smaller H$\beta$ EWs have higher metallicities at fixed electron density.

We present the electron densities, H$\beta$ EW, and N2 ratio (as a probe of metallicity) for galaxies in AURORA in Figure~\ref{fig:hbew}b.
For the AURORA sample, we consider the $\rm [NII]\lambda6584/H\alpha$ ratio as an indicator of the metallicity due to the monotonic relationship between the two quantities \citep[e.g.,][]{pp04, Marino2013}.
We present the electron densities of the AURORA sample as a function of H$\beta$ EW in Figure~\ref{fig:hbew}b, where each galaxy is identified by its $\rm [NII]/H\alpha$.
Although the AURORA sample spans a slightly smaller range of EWs, we find that many features of the simulated galaxies are represented in our observations.
Specifically, we find a positive correlation between H$\beta$ EW and electron density, but also a trend between density and $\rm [NII]/H\alpha$ ratio at fixed EW.
For galaxies with the same H$\beta$ EWs, objects with higher $\rm [NII]/H\alpha$ (i.e., higher metallicity) also have a higher electron density. 
Just as in the SPHINX simulations, this effect is most prominent at low H$\beta$ EWs, where the sample spans a much larger range of line ratio.
For just these objects ($\rm H\beta<40\angstrom{}$), we find that galaxies with low N2 ratios ($\log([\rm NII]/H\alpha)\le-0.7$) have an average density of $85~\rm cm^{-3}$, while galaxies with $\log([\rm NII]/H\alpha)\ge-0.5$ are described by an average density of $600~\rm cm^{-3}$, which is a factor of 7$\times$ higher.
Thus, it appears that at fixed metallicity there exists a significant correlation between H$\beta$ EW (or stellar age) and electron density,
although no such correlation is present when galaxies at all metallicities are  considered simultaneously.

These results qualitatively suggest that the simulated galaxies may be capturing key processes that drive the observed emission between galaxies.
Galaxies with the largest H$\beta$ EWs host the youngest ($\sim$Myrs) HII regions that may not yet have completely relaxed from their parent molecular cloud. Therefore, their densities may more closely reflect the dense conditions resulting from over-pressurization of the parent neutral gas from which their stars formed \citep[e.g.,][]{Oey1997, Nath2020}. 
The observed trend with metallicity at fixed H$\beta$ EW likely arises from a combination of effects. 
For HII regions in sustained equilibrium, the product of electron density and temperature ($n_e \times T_e$) is expected to remain constant. 
Consequently, gas at higher metallicities that will have lower electron temperatures may require higher densities to maintain this balance.
In addition, higher metallicity gas may also coincide with a higher metallicity stellar population, which produces a softer ionizing radiation field.
The weakened ionizing conditions may cause the part of the HII region that produces the line emission (e.g., for [SII]) to be closer to the ionizing source. 
If the density structure of the ISM is such that the densest gas exists closest to ionizing sources (see Section~\ref{sec:structure}), this may result in line emission that traces the gas at a higher density.
Finally, at fixed H$\beta$ EW, the trend between density and metallicity may also reflect some trend between density and stellar mass, due to the relation between stellar mass and metallicity \citep[MZR; e.g.,][]{Lequeux1979, Tremonti2004, Lee2006, Erb2006, Kewley2008, Berg2012, Andrews2013, Sanders2021, Curti2024}. 
Collectively, these findings demonstrate that the electron densities are closely interconnected with the properties of gas and stars within galaxies and that these connections can be described by galaxy formation simulations.
With the increased quality of high-redshift galaxy spectra from {\it JWST}, the interpretation of galaxy spectra will continue to benefit from more realistic simulations that can be used to self-consistently model the internal properties of high-redshift galaxies.

%
%
%
%
\section{Summary}
\label{sec:summary}

In this paper, we have analyzed deep {\it JWST}/NIRSpec spectra of star-forming galaxies spanning $z\simeq1.4-10$ observed as part of the AURORA survey.
We summarize our main results below.

(i) We inferred electron densities for a sample of galaxies spanning redshifts of $z\simeq1.4-6$, and found densities from [SII] of $268^{+45}_{-49}~\rm cm^{-3}$, $350^{+140}_{-76}~\rm cm^{-3}$, and $480^{+390}_{-310}~\rm cm^{-3}$ at redshifts of $z=2.3$, $z=3.2$, and $z=5.3$.
These medians represent an evolutionary trend of electron densities that increases toward a higher redshift at a rate of $(1+z)^{1.5\pm0.6}$.
We find that this trend derived for galaxies at $z\simeq1.4-6$ is consistent with galaxies extending beyond this range from $z\simeq0-10$ from the literature, and that it evolves slower than what we expect based on galaxy size evolution at fixed stellar mass.

(ii) We compared the electron densities of AURORA galaxies to their integrated galaxy properties. 
We find relatively weak positive correlations between density and SFR ($\log(\rm n_e/cm^{-3})=(0.29\pm 0.16)\times\log(\frac{\rm SFR}{M_{\odot}yr^{-1}})+(2.18\pm 0.13)$) and between density and SFR surface density ($\log(\rm n_e/cm^{-3})=(0.15\pm0.10)\times\log(\frac{\Sigma_{\rm \rm SFR}}{M_{\odot}yr^{-1}kpc^{-2}})+(2.42\pm0.08)$), and no significant trends between density and stellar mass or sSFR.

(iii) We compared the electron densities of the AURORA sample with rest-optical emission line ratios and found that densities become larger with increasing Ne3O2. We suggest that a similar trend may exist between density and O32, but may not be detected in our sample due to galaxy-to-galaxy variations in the dust attenuation curve (and associated emission line correction). Furthermore, while neither [OIII]/H$\beta$ or [NII]/H$\alpha$ is correlated with electron density on its own, we found that density increases with distance away from the local BPT sequence. The density increase is most strongly correlated with perpendicular distance from the local BPT locus, and the rate of change in the density with BPT location is greater than what is expected from photoionization models alone.

(iv)  We derive densities from the CIII] rest-frame UV doublet for AURORA galaxies $z>4$ to probe the density in a higher-ionization phase of the ISM. 
We find a median density of $1.4^{+0.7}_{-0.5}\times10^4~\rm cm^{-3}$, which is a factor of $30$ higher on average compared to the [SII] densities in the same redshift range. 
We consider additional CIII] measurements at lower redshift from the literature, from which we find a redshift evolution of the densities of $\propto(1+z)^{1.6}$, which is consistent with the rate derived for [SII]. 
This co-evolution of [SII] and CIII] densities may indicate that the structure of HII regions is consistent throughout cosmic time, and is such that the interiors are composed of more dense, higher-ionization gas compared to the less dense, low-ionization exteriors.

(v) We compared galaxies in the AURORA sample to galaxies calculated from the SPHINX cosmological radiation-hydrodynamic simulations that make predictions of the emission from multiple phases of density within the ISM.
We used the H$\beta$ equivalent widths as an empirical probe for galaxy age for the same region of the galaxy from which the electron densities were derived.
In both the simulations and observational data sets, we find that galaxies form a multi-dimensional relation between H$\beta$ EW, density, and metallicity (parameterized by [NII]/H$\alpha$ for AURORA). These trends are such that galaxies with larger H$\beta$ EWs typically have higher electron densities, with a secondary positive correlation between density and metallicity at fixed H$\beta$ EW.

The results presented here indicate that the ISM of high-redshift galaxies is not composed of a single, monolithic density component. Instead, gas that spans multiple orders-of-magnitude in electron density is responsible for their observed emission from galaxies spanning cosmic time.
This notion must be considered when interpreting galaxy spectra self consistently, and in doing so we will continue to improve our understanding of early galaxy properties in the era of {\it JWST}.

\section*{Acknowledgements}
This work is based in part on observations made with the NASA/ESA/CSA James Webb Space Telescope. The data were obtained from the Mikulski Archive for Space Telescopes at the Space Telescope Science Institute, which is operated by the Association of Universities for Research in Astronomy, Inc., under NASA contract NAS 5-03127 for {\it JWST}. These observations are associated with GO program ID 1914.
T.J. acknowledges support from the National Aeronautics and Space Administration
(NASA) under grants 80NSSC23K1132 and HST-GO-16697, and from a Chancellor's Fellowship.
JSD acknowledges the support of the Royal Society via the award of a Royal Society Research Professorship.

 \section*{Data Availability}
The data underlying this article may be produced upon reasonable request to the corresponding author.

\bibliographystyle{mnras}
\bibliography{main}

\begin{thebibliography}{}
\makeatletter
\relax
\def\mn@urlcharsother{\let\do\@makeother \do\$\do\&\do\#\do\^\do\_\do\%\do\~}
\def\mn@doi{\begingroup\mn@urlcharsother \@ifnextchar [ {\mn@doi@} {\mn@doi@[]}}
\def\mn@doi@[#1]#2{\def\@tempa{#1}\ifx\@tempa\@empty \href {http://dx.doi.org/#2} {doi:#2}\else \href {http://dx.doi.org/#2} {#1}\fi \endgroup}
\def\mn@eprint#1#2{\mn@eprint@#1:#2::\@nil}
\def\mn@eprint@arXiv#1{\href {http://arxiv.org/abs/#1} {{\tt arXiv:#1}}}
\def\mn@eprint@dblp#1{\href {http://dblp.uni-trier.de/rec/bibtex/#1.xml} {dblp:#1}}
\def\mn@eprint@#1:#2:#3:#4\@nil{\def\@tempa {#1}\def\@tempb {#2}\def\@tempc {#3}\ifx \@tempc \@empty \let \@tempc \@tempb \let \@tempb \@tempa \fi \ifx \@tempb \@empty \def\@tempb {arXiv}\fi \@ifundefined {mn@eprint@\@tempb}{\@tempb:\@tempc}{\expandafter \expandafter \csname mn@eprint@\@tempb\endcsname \expandafter{\@tempc}}}

\bibitem[\protect\citeauthoryear{{Abdurro'uf} et~al.,}{{Abdurro'uf} et~al.}{2024}]{Abdurrouf2024}
{Abdurro'uf} et~al., 2024, \mn@doi [arXiv e-prints] {10.48550/arXiv.2404.16201}, \href {https://ui.adsabs.harvard.edu/abs/2024arXiv240416201A} {p. arXiv:2404.16201}

\bibitem[\protect\citeauthoryear{{Andrews} \& {Martini}}{{Andrews} \& {Martini}}{2013}]{Andrews2013}
{Andrews} B.~H.,  {Martini} P.,  2013, \mn@doi [\apj] {10.1088/0004-637X/765/2/140}, \href {https://ui.adsabs.harvard.edu/abs/2013ApJ...765..140A} {765, 140}

\bibitem[\protect\citeauthoryear{{Asplund}, {Grevesse}, {Sauval}  \& {Scott}}{{Asplund} et~al.}{2009}]{Asplund2009}
{Asplund} M.,  {Grevesse} N.,  {Sauval} A.~J.,   {Scott} P.,  2009, \mn@doi [\araa] {10.1146/annurev.astro.46.060407.145222}, \href {https://ui.adsabs.harvard.edu/abs/2009ARA&A..47..481A} {47, 481}

\bibitem[\protect\citeauthoryear{{Baldwin}, {Phillips}  \& {Terlevich}}{{Baldwin} et~al.}{1981}]{Baldwin1981}
{Baldwin} J.~A.,  {Phillips} M.~M.,   {Terlevich} R.,  1981, \mn@doi [\pasp] {10.1086/130766}, \href {https://ui.adsabs.harvard.edu/abs/1981PASP...93....5B} {93, 5}

\bibitem[\protect\citeauthoryear{{Berg} et~al.,}{{Berg} et~al.}{2012}]{Berg2012}
{Berg} D.~A.,  et~al., 2012, \mn@doi [\apj] {10.1088/0004-637X/754/2/98}, \href {https://ui.adsabs.harvard.edu/abs/2012ApJ...754...98B} {754, 98}

\bibitem[\protect\citeauthoryear{{Berg}, {Chisholm}, {Erb}, {Skillman}, {Pogge}  \& {Olivier}}{{Berg} et~al.}{2021}]{Berg2021}
{Berg} D.~A.,  {Chisholm} J.,  {Erb} D.~K.,  {Skillman} E.~D.,  {Pogge} R.~W.,   {Olivier} G.~M.,  2021, \mn@doi [\apj] {10.3847/1538-4357/ac141b}, \href {https://ui.adsabs.harvard.edu/abs/2021ApJ...922..170B} {922, 170}

\bibitem[\protect\citeauthoryear{{Berg} et~al.,}{{Berg} et~al.}{2022}]{Berg2022}
{Berg} D.~A.,  et~al., 2022, \mn@doi [\apjs] {10.3847/1538-4365/ac6c03}, \href {https://ui.adsabs.harvard.edu/abs/2022ApJS..261...31B} {261, 31}

\bibitem[\protect\citeauthoryear{{Bian} et~al.,}{{Bian} et~al.}{2010}]{Bian2010}
{Bian} F.,  et~al., 2010, \mn@doi [\apj] {10.1088/0004-637X/725/2/1877}, \href {https://ui.adsabs.harvard.edu/abs/2010ApJ...725.1877B} {725, 1877}

\bibitem[\protect\citeauthoryear{{Bian}, {Kewley}, {Dopita}  \& {Juneau}}{{Bian} et~al.}{2016}]{Bian2016}
{Bian} F.,  {Kewley} L.~J.,  {Dopita} M.~A.,   {Juneau} S.,  2016, \mn@doi [\apj] {10.3847/0004-637X/822/2/62}, \href {https://ui.adsabs.harvard.edu/abs/2016ApJ...822...62B} {822, 62}

\bibitem[\protect\citeauthoryear{{Bouwens} et~al.,}{{Bouwens} et~al.}{2015}]{Bouwens2015}
{Bouwens} R.~J.,  et~al., 2015, \mn@doi [\apj] {10.1088/0004-637X/803/1/34}, \href {https://ui.adsabs.harvard.edu/abs/2015ApJ...803...34B} {803, 34}

\bibitem[\protect\citeauthoryear{{Brinchmann}, {Pettini}  \& {Charlot}}{{Brinchmann} et~al.}{2008}]{Brinchmann2008}
{Brinchmann} J.,  {Pettini} M.,   {Charlot} S.,  2008, \mn@doi [\mnras] {10.1111/j.1365-2966.2008.12914.x}, \href {https://ui.adsabs.harvard.edu/abs/2008MNRAS.385..769B} {385, 769}

\bibitem[\protect\citeauthoryear{{Bunker} et~al.,}{{Bunker} et~al.}{2023}]{Bunker2023}
{Bunker} A.~J.,  et~al., 2023, \mn@doi [\aap] {10.1051/0004-6361/202346159}, \href {https://ui.adsabs.harvard.edu/abs/2023A&A...677A..88B} {677, A88}

\bibitem[\protect\citeauthoryear{{Byler}, {Dalcanton}, {Conroy}  \& {Johnson}}{{Byler} et~al.}{2017}]{Byler2017}
{Byler} N.,  {Dalcanton} J.~J.,  {Conroy} C.,   {Johnson} B.~D.,  2017, \mn@doi [\apj] {10.3847/1538-4357/aa6c66}, \href {https://ui.adsabs.harvard.edu/abs/2017ApJ...840...44B} {840, 44}

\bibitem[\protect\citeauthoryear{{Calzetti}, {Armus}, {Bohlin}, {Kinney}, {Koornneef}  \& {Storchi-Bergmann}}{{Calzetti} et~al.}{2000}]{Calzetti2000}
{Calzetti} D.,  {Armus} L.,  {Bohlin} R.~C.,  {Kinney} A.~L.,  {Koornneef} J.,   {Storchi-Bergmann} T.,  2000, \mn@doi [\apj] {10.1086/308692}, \href {https://ui.adsabs.harvard.edu/abs/2000ApJ...533..682C} {533, 682}

\bibitem[\protect\citeauthoryear{{Cameron}, {Katz}, {Witten}, {Saxena}, {Laporte}  \& {Bunker}}{{Cameron} et~al.}{2024}]{Cameron2023}
{Cameron} A.~J.,  {Katz} H.,  {Witten} C.,  {Saxena} A.,  {Laporte} N.,   {Bunker} A.~J.,  2024, \mn@doi [\mnras] {10.1093/mnras/stae1547}, \href {https://ui.adsabs.harvard.edu/abs/2024MNRAS.534..523C} {534, 523}

\bibitem[\protect\citeauthoryear{{Cardelli}, {Clayton}  \& {Mathis}}{{Cardelli} et~al.}{1989}]{Cardelli1989}
{Cardelli} J.~A.,  {Clayton} G.~C.,   {Mathis} J.~S.,  1989, \mn@doi [\apj] {10.1086/167900}, \href {https://ui.adsabs.harvard.edu/abs/1989ApJ...345..245C} {345, 245}

\bibitem[\protect\citeauthoryear{{Chabrier}}{{Chabrier}}{2003}]{Chabrier2003}
{Chabrier} G.,  2003, \mn@doi [\pasp] {10.1086/376392}, \href {https://ui.adsabs.harvard.edu/abs/2003PASP..115..763C} {115, 763}

\bibitem[\protect\citeauthoryear{{Conroy} \& {Gunn}}{{Conroy} \& {Gunn}}{2010}]{Conroy2010}
{Conroy} C.,  {Gunn} J.~E.,  2010, \mn@doi [\apj] {10.1088/0004-637X/712/2/833}, \href {https://ui.adsabs.harvard.edu/abs/2010ApJ...712..833C} {712, 833}

\bibitem[\protect\citeauthoryear{{Conroy}, {Gunn}  \& {White}}{{Conroy} et~al.}{2009}]{Conroy2009}
{Conroy} C.,  {Gunn} J.~E.,   {White} M.,  2009, \mn@doi [\apj] {10.1088/0004-637X/699/1/486}, \href {https://ui.adsabs.harvard.edu/abs/2009ApJ...699..486C} {699, 486}

\bibitem[\protect\citeauthoryear{{Curti} et~al.,}{{Curti} et~al.}{2024}]{Curti2024}
{Curti} M.,  et~al., 2024, \mn@doi [\aap] {10.1051/0004-6361/202346698}, \href {https://ui.adsabs.harvard.edu/abs/2024A&A...684A..75C} {684, A75}

\bibitem[\protect\citeauthoryear{{Davies} et~al.,}{{Davies} et~al.}{2021}]{Davies2021}
{Davies} R.~L.,  et~al., 2021, \mn@doi [\apj] {10.3847/1538-4357/abd551}, \href {https://ui.adsabs.harvard.edu/abs/2021ApJ...909...78D} {909, 78}

\bibitem[\protect\citeauthoryear{{Dunlop} et~al.,}{{Dunlop} et~al.}{2021}]{Dunlop2021}
{Dunlop} J.~S.,  et~al., 2021, {PRIMER: Public Release IMaging for Extragalactic Research}, JWST Proposal. Cycle 1, ID. \#1837

\bibitem[\protect\citeauthoryear{{Eisenstein} et~al.,}{{Eisenstein} et~al.}{2023}]{Eisenstein2023}
{Eisenstein} D.~J.,  et~al., 2023, \mn@doi [arXiv e-prints] {10.48550/arXiv.2306.02465}, \href {https://ui.adsabs.harvard.edu/abs/2023arXiv230602465E} {p. arXiv:2306.02465}

\bibitem[\protect\citeauthoryear{{Endsley} et~al.,}{{Endsley} et~al.}{2024}]{Endsley2024}
{Endsley} R.,  et~al., 2024, \mn@doi [\mnras] {10.1093/mnras/stae1857}, \href {https://ui.adsabs.harvard.edu/abs/2024MNRAS.533.1111E} {533, 1111}

\bibitem[\protect\citeauthoryear{{Erb}, {Shapley}, {Pettini}, {Steidel}, {Reddy}  \& {Adelberger}}{{Erb} et~al.}{2006}]{Erb2006}
{Erb} D.~K.,  {Shapley} A.~E.,  {Pettini} M.,  {Steidel} C.~C.,  {Reddy} N.~A.,   {Adelberger} K.~L.,  2006, \mn@doi [\apj] {10.1086/503623}, \href {https://ui.adsabs.harvard.edu/abs/2006ApJ...644..813E} {644, 813}

\bibitem[\protect\citeauthoryear{{Finkelstein} et~al.,}{{Finkelstein} et~al.}{2015}]{Finkelstein2015}
{Finkelstein} S.~L.,  et~al., 2015, \mn@doi [\apj] {10.1088/0004-637X/810/1/71}, \href {https://ui.adsabs.harvard.edu/abs/2015ApJ...810...71F} {810, 71}

\bibitem[\protect\citeauthoryear{{Fujimoto} et~al.,}{{Fujimoto} et~al.}{2023}]{Fujimoto2023}
{Fujimoto} S.,  et~al., 2023, \mn@doi [\apjl] {10.3847/2041-8213/acd2d9}, \href {https://ui.adsabs.harvard.edu/abs/2023ApJ...949L..25F} {949, L25}

\bibitem[\protect\citeauthoryear{{Gordon}, {Clayton}, {Misselt}, {Landolt}  \& {Wolff}}{{Gordon} et~al.}{2003}]{Gordon2003}
{Gordon} K.~D.,  {Clayton} G.~C.,  {Misselt} K.~A.,  {Landolt} A.~U.,   {Wolff} M.~J.,  2003, \mn@doi [\apj] {10.1086/376774}, \href {https://ui.adsabs.harvard.edu/abs/2003ApJ...594..279G} {594, 279}

\bibitem[\protect\citeauthoryear{{Grogin} et~al.,}{{Grogin} et~al.}{2011}]{Grogin2011}
{Grogin} N.~A.,  et~al., 2011, \mn@doi [\apjs] {10.1088/0067-0049/197/2/35}, \href {https://ui.adsabs.harvard.edu/abs/2011ApJS..197...35G} {197, 35}

\bibitem[\protect\citeauthoryear{{Hainline}, {Shapley}, {Kornei}, {Pettini}, {Buckley-Geer}, {Allam}  \& {Tucker}}{{Hainline} et~al.}{2009}]{Hainline2009}
{Hainline} K.~N.,  {Shapley} A.~E.,  {Kornei} K.~A.,  {Pettini} M.,  {Buckley-Geer} E.,  {Allam} S.~S.,   {Tucker} D.~L.,  2009, \mn@doi [\apj] {10.1088/0004-637X/701/1/52}, \href {https://ui.adsabs.harvard.edu/abs/2009ApJ...701...52H} {701, 52}

\bibitem[\protect\citeauthoryear{{Hainline} et~al.,}{{Hainline} et~al.}{2024}]{Hainline2024}
{Hainline} K.~N.,  et~al., 2024, \mn@doi [\apj] {10.3847/1538-4357/ad1ee4}, \href {https://ui.adsabs.harvard.edu/abs/2024ApJ...964...71H} {964, 71}

\bibitem[\protect\citeauthoryear{{Harshan} et~al.,}{{Harshan} et~al.}{2020}]{Harshan2020}
{Harshan} A.,  et~al., 2020, \mn@doi [\apj] {10.3847/1538-4357/ab76cf}, \href {https://ui.adsabs.harvard.edu/abs/2020ApJ...892...77H} {892, 77}

\bibitem[\protect\citeauthoryear{{Heintz} et~al.,}{{Heintz} et~al.}{2024}]{Heintz2024}
{Heintz} K.~E.,  et~al., 2024, \mn@doi [arXiv e-prints] {10.48550/arXiv.2404.02211}, \href {https://ui.adsabs.harvard.edu/abs/2024arXiv240402211H} {p. arXiv:2404.02211}

\bibitem[\protect\citeauthoryear{{Herrera-Camus} et~al.,}{{Herrera-Camus} et~al.}{2016}]{HerreraCamus2016}
{Herrera-Camus} R.,  et~al., 2016, \mn@doi [\apj] {10.3847/0004-637X/826/2/175}, \href {https://ui.adsabs.harvard.edu/abs/2016ApJ...826..175H} {826, 175}

\bibitem[\protect\citeauthoryear{{Horne}}{{Horne}}{1986}]{Horne1986}
{Horne} K.,  1986, \mn@doi [\pasp] {10.1086/131801}, \href {https://ui.adsabs.harvard.edu/abs/1986PASP...98..609H} {98, 609}

\bibitem[\protect\citeauthoryear{{Hu} et~al.,}{{Hu} et~al.}{2024}]{Hu2024}
{Hu} W.,  et~al., 2024, \mn@doi [\apj] {10.3847/1538-4357/ad5015}, \href {https://ui.adsabs.harvard.edu/abs/2024ApJ...971...21H} {971, 21}

\bibitem[\protect\citeauthoryear{{Hunt} \& {Hirashita}}{{Hunt} \& {Hirashita}}{2009}]{Hunt2009}
{Hunt} L.~K.,  {Hirashita} H.,  2009, \mn@doi [\aap] {10.1051/0004-6361/200912020}, \href {https://ui.adsabs.harvard.edu/abs/2009A&A...507.1327H} {507, 1327}

\bibitem[\protect\citeauthoryear{{Isobe}, {Ouchi}, {Nakajima}, {Harikane}, {Ono}, {Xu}, {Zhang}  \& {Umeda}}{{Isobe} et~al.}{2023}]{Isobe2023}
{Isobe} Y.,  {Ouchi} M.,  {Nakajima} K.,  {Harikane} Y.,  {Ono} Y.,  {Xu} Y.,  {Zhang} Y.,   {Umeda} H.,  2023, \mn@doi [\apj] {10.3847/1538-4357/acf376}, \href {https://ui.adsabs.harvard.edu/abs/2023ApJ...956..139I} {956, 139}

\bibitem[\protect\citeauthoryear{{James} et~al.,}{{James} et~al.}{2014}]{James2014}
{James} B.~L.,  et~al., 2014, \mn@doi [\mnras] {10.1093/mnras/stu287}, \href {https://ui.adsabs.harvard.edu/abs/2014MNRAS.440.1794J} {440, 1794}

\bibitem[\protect\citeauthoryear{{Jiang}, {Dekel}, {Freundlich}, {Romanowsky}, {Dutton}, {Macci{\`o}}  \& {Di Cintio}}{{Jiang} et~al.}{2019}]{Jiang2019}
{Jiang} F.,  {Dekel} A.,  {Freundlich} J.,  {Romanowsky} A.~J.,  {Dutton} A.~A.,  {Macci{\`o}} A.~V.,   {Di Cintio} A.,  2019, \mn@doi [\mnras] {10.1093/mnras/stz1499}, \href {https://ui.adsabs.harvard.edu/abs/2019MNRAS.487.5272J} {487, 5272}

\bibitem[\protect\citeauthoryear{{Johnson}, {Leja}, {Conroy}  \& {Speagle}}{{Johnson} et~al.}{2021}]{Johnson2021}
{Johnson} B.~D.,  {Leja} J.,  {Conroy} C.,   {Speagle} J.~S.,  2021, \mn@doi [\apjs] {10.3847/1538-4365/abef67}, \href {https://ui.adsabs.harvard.edu/abs/2021ApJS..254...22J} {254, 22}

\bibitem[\protect\citeauthoryear{{Jung} et~al.,}{{Jung} et~al.}{2020}]{Jung2020}
{Jung} I.,  et~al., 2020, \mn@doi [\apj] {10.3847/1538-4357/abbd44}, \href {https://ui.adsabs.harvard.edu/abs/2020ApJ...904..144J} {904, 144}

\bibitem[\protect\citeauthoryear{{Kaasinen}, {Bian}, {Groves}, {Kewley}  \& {Gupta}}{{Kaasinen} et~al.}{2017}]{Kaasinen2017}
{Kaasinen} M.,  {Bian} F.,  {Groves} B.,  {Kewley} L.~J.,   {Gupta} A.,  2017, \mn@doi [\mnras] {10.1093/mnras/stw2827}, \href {https://ui.adsabs.harvard.edu/abs/2017MNRAS.465.3220K} {465, 3220}

\bibitem[\protect\citeauthoryear{{Kashino} \& {Inoue}}{{Kashino} \& {Inoue}}{2019}]{Kashino2019}
{Kashino} D.,  {Inoue} A.~K.,  2019, \mn@doi [\mnras] {10.1093/mnras/stz881}, \href {https://ui.adsabs.harvard.edu/abs/2019MNRAS.486.1053K} {486, 1053}

\bibitem[\protect\citeauthoryear{{Kashino} et~al.,}{{Kashino} et~al.}{2017}]{Kashino2017}
{Kashino} D.,  et~al., 2017, \mn@doi [\apj] {10.3847/1538-4357/835/1/88}, \href {https://ui.adsabs.harvard.edu/abs/2017ApJ...835...88K} {835, 88}

\bibitem[\protect\citeauthoryear{{Katz} et~al.,}{{Katz} et~al.}{2023}]{Katz2023}
{Katz} H.,  et~al., 2023, \mn@doi [The Open Journal of Astrophysics] {10.21105/astro.2309.03269}, \href {https://ui.adsabs.harvard.edu/abs/2023OJAp....6E..44K} {6, 44}

\bibitem[\protect\citeauthoryear{{Katz} et~al.,}{{Katz} et~al.}{2024}]{Katz2024}
{Katz} H.,  et~al., 2024, \mn@doi [arXiv e-prints] {10.48550/arXiv.2408.03189}, \href {https://ui.adsabs.harvard.edu/abs/2024arXiv240803189K} {p. arXiv:2408.03189}

\bibitem[\protect\citeauthoryear{{Kelly}}{{Kelly}}{2007}]{Kelly2007}
{Kelly} B.~C.,  2007, \mn@doi [\apj] {10.1086/519947}, \href {https://ui.adsabs.harvard.edu/abs/2007ApJ...665.1489K} {665, 1489}

\bibitem[\protect\citeauthoryear{{Kewley} \& {Ellison}}{{Kewley} \& {Ellison}}{2008}]{Kewley2008}
{Kewley} L.~J.,  {Ellison} S.~L.,  2008, \mn@doi [\apj] {10.1086/587500}, \href {https://ui.adsabs.harvard.edu/abs/2008ApJ...681.1183K} {681, 1183}

\bibitem[\protect\citeauthoryear{{Kewley}, {Dopita}, {Leitherer}, {Dav{\'e}}, {Yuan}, {Allen}, {Groves}  \& {Sutherland}}{{Kewley} et~al.}{2013}]{Kewley2013}
{Kewley} L.~J.,  {Dopita} M.~A.,  {Leitherer} C.,  {Dav{\'e}} R.,  {Yuan} T.,  {Allen} M.,  {Groves} B.,   {Sutherland} R.,  2013, \mn@doi [\apj] {10.1088/0004-637X/774/2/100}, \href {https://ui.adsabs.harvard.edu/abs/2013ApJ...774..100K} {774, 100}

\bibitem[\protect\citeauthoryear{{Kriek}, {van Dokkum}, {Labb{\'e}}, {Franx}, {Illingworth}, {Marchesini}  \& {Quadri}}{{Kriek} et~al.}{2009}]{Kriek2009}
{Kriek} M.,  {van Dokkum} P.~G.,  {Labb{\'e}} I.,  {Franx} M.,  {Illingworth} G.~D.,  {Marchesini} D.,   {Quadri} R.~F.,  2009, \mn@doi [\apj] {10.1088/0004-637X/700/1/221}, \href {https://ui.adsabs.harvard.edu/abs/2009ApJ...700..221K} {700, 221}

\bibitem[\protect\citeauthoryear{{Kumari} et~al.,}{{Kumari} et~al.}{2024}]{Kumari2024}
{Kumari} N.,  et~al., 2024, \mn@doi [arXiv e-prints] {10.48550/arXiv.2406.11997}, \href {https://ui.adsabs.harvard.edu/abs/2024arXiv240611997K} {p. arXiv:2406.11997}

\bibitem[\protect\citeauthoryear{{Laseter} et~al.,}{{Laseter} et~al.}{2024}]{Laseter2024}
{Laseter} I.~H.,  et~al., 2024, \mn@doi [\aap] {10.1051/0004-6361/202347133}, \href {https://ui.adsabs.harvard.edu/abs/2024A&A...681A..70L} {681, A70}

\bibitem[\protect\citeauthoryear{{Lee}, {Skillman}, {Cannon}, {Jackson}, {Gehrz}, {Polomski}  \& {Woodward}}{{Lee} et~al.}{2006}]{Lee2006}
{Lee} H.,  {Skillman} E.~D.,  {Cannon} J.~M.,  {Jackson} D.~C.,  {Gehrz} R.~D.,  {Polomski} E.~F.,   {Woodward} C.~E.,  2006, \mn@doi [\apj] {10.1086/505573}, \href {https://ui.adsabs.harvard.edu/abs/2006ApJ...647..970L} {647, 970}

\bibitem[\protect\citeauthoryear{{Lehnert}, {Nesvadba}, {Le Tiran}, {Di Matteo}, {van Driel}, {Douglas}, {Chemin}  \& {Bournaud}}{{Lehnert} et~al.}{2009}]{Lehnert2009}
{Lehnert} M.~D.,  {Nesvadba} N.~P.~H.,  {Le Tiran} L.,  {Di Matteo} P.,  {van Driel} W.,  {Douglas} L.~S.,  {Chemin} L.,   {Bournaud} F.,  2009, \mn@doi [\apj] {10.1088/0004-637X/699/2/1660}, \href {https://ui.adsabs.harvard.edu/abs/2009ApJ...699.1660L} {699, 1660}

\bibitem[\protect\citeauthoryear{{Lequeux}, {Peimbert}, {Rayo}, {Serrano}  \& {Torres-Peimbert}}{{Lequeux} et~al.}{1979}]{Lequeux1979}
{Lequeux} J.,  {Peimbert} M.,  {Rayo} J.~F.,  {Serrano} A.,   {Torres-Peimbert} S.,  1979, \aap, \href {https://ui.adsabs.harvard.edu/abs/1979A&A....80..155L} {80, 155}

\bibitem[\protect\citeauthoryear{{Li} et~al.,}{{Li} et~al.}{2024}]{Li2024}
{Li} S.,  et~al., 2024, \mn@doi [arXiv e-prints] {10.48550/arXiv.2412.08382}, \href {https://ui.adsabs.harvard.edu/abs/2024arXiv241208382L} {p. arXiv:2412.08382}

\bibitem[\protect\citeauthoryear{{Luridiana}, {Morisset}  \& {Shaw}}{{Luridiana} et~al.}{2015}]{Luridiana2015}
{Luridiana} V.,  {Morisset} C.,   {Shaw} R.~A.,  2015, \mn@doi [\aap] {10.1051/0004-6361/201323152}, \href {https://ui.adsabs.harvard.edu/abs/2015A&A...573A..42L} {573, A42}

\bibitem[\protect\citeauthoryear{{Mainali}, {Stark}, {Jones}, {Ellis}, {Hezaveh}  \& {Rigby}}{{Mainali} et~al.}{2023}]{Mainali2023}
{Mainali} R.,  {Stark} D.~P.,  {Jones} T.,  {Ellis} R.~S.,  {Hezaveh} Y.~D.,   {Rigby} J.~R.,  2023, \mn@doi [\mnras] {10.1093/mnras/stad387}, \href {https://ui.adsabs.harvard.edu/abs/2023MNRAS.520.4037M} {520, 4037}

\bibitem[\protect\citeauthoryear{{Marino} et~al.,}{{Marino} et~al.}{2013}]{Marino2013}
{Marino} R.~A.,  et~al., 2013, \mn@doi [\aap] {10.1051/0004-6361/201321956}, \href {https://ui.adsabs.harvard.edu/abs/2013A&A...559A.114M} {559, A114}

\bibitem[\protect\citeauthoryear{{Martorano}, {van der Wel}, {Baes}, {Bell}, {Brammer}, {Franx}  \& {Nersesian}}{{Martorano} et~al.}{2024}]{Martorano2024}
{Martorano} M.,  {van der Wel} A.,  {Baes} M.,  {Bell} E.~F.,  {Brammer} G.,  {Franx} M.,   {Nersesian} A.,  2024, \mn@doi [\apj] {10.3847/1538-4357/ad5c6a}, \href {https://ui.adsabs.harvard.edu/abs/2024ApJ...972..134M} {972, 134}

\bibitem[\protect\citeauthoryear{{Maseda} et~al.,}{{Maseda} et~al.}{2017}]{Maseda2017}
{Maseda} M.~V.,  et~al., 2017, \mn@doi [\aap] {10.1051/0004-6361/201730985}, \href {https://ui.adsabs.harvard.edu/abs/2017A&A...608A...4M} {608, A4}

\bibitem[\protect\citeauthoryear{{Mingozzi} et~al.,}{{Mingozzi} et~al.}{2022}]{Mingozzi2022}
{Mingozzi} M.,  et~al., 2022, \mn@doi [\apj] {10.3847/1538-4357/ac952c}, \href {https://ui.adsabs.harvard.edu/abs/2022ApJ...939..110M} {939, 110}

\bibitem[\protect\citeauthoryear{{Nakajima}, {Ouchi}, {Isobe}, {Harikane}, {Zhang}, {Ono}, {Umeda}  \& {Oguri}}{{Nakajima} et~al.}{2023}]{Nakajima2023}
{Nakajima} K.,  {Ouchi} M.,  {Isobe} Y.,  {Harikane} Y.,  {Zhang} Y.,  {Ono} Y.,  {Umeda} H.,   {Oguri} M.,  2023, \mn@doi [\apjs] {10.3847/1538-4365/acd556}, \href {https://ui.adsabs.harvard.edu/abs/2023ApJS..269...33N} {269, 33}

\bibitem[\protect\citeauthoryear{{Nath}, {Das}  \& {Oey}}{{Nath} et~al.}{2020}]{Nath2020}
{Nath} B.~B.,  {Das} P.,   {Oey} M.~S.,  2020, \mn@doi [\mnras] {10.1093/mnras/staa336}, \href {https://ui.adsabs.harvard.edu/abs/2020MNRAS.493.1034N} {493, 1034}

\bibitem[\protect\citeauthoryear{{Oesch} et~al.,}{{Oesch} et~al.}{2023}]{Oesch2023}
{Oesch} P.~A.,  et~al., 2023, \mn@doi [\mnras] {10.1093/mnras/stad2411}, \href {https://ui.adsabs.harvard.edu/abs/2023MNRAS.525.2864O} {525, 2864}

\bibitem[\protect\citeauthoryear{{Oey} \& {Clarke}}{{Oey} \& {Clarke}}{1997}]{Oey1997}
{Oey} M.~S.,  {Clarke} C.~J.,  1997, \mn@doi [\mnras] {10.1093/mnras/289.3.570}, \href {https://ui.adsabs.harvard.edu/abs/1997MNRAS.289..570O} {289, 570}

\bibitem[\protect\citeauthoryear{{Oke} \& {Gunn}}{{Oke} \& {Gunn}}{1983}]{Oke1984}
{Oke} J.~B.,  {Gunn} J.~E.,  1983, \mn@doi [\apj] {10.1086/160817}, \href {https://ui.adsabs.harvard.edu/abs/1983ApJ...266..713O} {266, 713}

\bibitem[\protect\citeauthoryear{{Osterbrock}}{{Osterbrock}}{1989}]{Osterbrock1989}
{Osterbrock} D.~E.,  1989, {Astrophysics of gaseous nebulae and active galactic nuclei}

\bibitem[\protect\citeauthoryear{{Osterbrock} \& {Ferland}}{{Osterbrock} \& {Ferland}}{2006}]{Osterbrock2006}
{Osterbrock} D.~E.,  {Ferland} G.~J.,  2006, {Astrophysics of gaseous nebulae and active galactic nuclei}

\bibitem[\protect\citeauthoryear{{Pahl}, {Shapley}, {Steidel}, {Reddy}  \& {Chen}}{{Pahl} et~al.}{2022}]{Pahl2022}
{Pahl} A.~J.,  {Shapley} A.,  {Steidel} C.~C.,  {Reddy} N.~A.,   {Chen} Y.,  2022, \mn@doi [\mnras] {10.1093/mnras/stac1767}, \href {https://ui.adsabs.harvard.edu/abs/2022MNRAS.516.2062P} {516, 2062}

\bibitem[\protect\citeauthoryear{{Peng}, {Ho}, {Impey}  \& {Rix}}{{Peng} et~al.}{2002}]{Peng2002}
{Peng} C.~Y.,  {Ho} L.~C.,  {Impey} C.~D.,   {Rix} H.-W.,  2002, \mn@doi [\aj] {10.1086/340952}, \href {https://ui.adsabs.harvard.edu/abs/2002AJ....124..266P} {124, 266}

\bibitem[\protect\citeauthoryear{{Peng}, {Ho}, {Impey}  \& {Rix}}{{Peng} et~al.}{2010}]{Peng2010}
{Peng} C.~Y.,  {Ho} L.~C.,  {Impey} C.~D.,   {Rix} H.-W.,  2010, \mn@doi [\aj] {10.1088/0004-6256/139/6/2097}, \href {https://ui.adsabs.harvard.edu/abs/2010AJ....139.2097P} {139, 2097}

\bibitem[\protect\citeauthoryear{{Pettini} \& {Pagel}}{{Pettini} \& {Pagel}}{2004}]{pp04}
{Pettini} M.,  {Pagel} B. E.~J.,  2004, \mn@doi [\mnras] {10.1111/j.1365-2966.2004.07591.x}, \href {https://ui.adsabs.harvard.edu/abs/2004MNRAS.348L..59P} {348, L59}

\bibitem[\protect\citeauthoryear{{Popesso} et~al.,}{{Popesso} et~al.}{2023}]{Popesso2023}
{Popesso} P.,  et~al., 2023, \mn@doi [\mnras] {10.1093/mnras/stac3214}, \href {https://ui.adsabs.harvard.edu/abs/2023MNRAS.519.1526P} {519, 1526}

\bibitem[\protect\citeauthoryear{{Reddy} et~al.,}{{Reddy} et~al.}{2022}]{Reddy2022}
{Reddy} N.~A.,  et~al., 2022, \mn@doi [\apj] {10.3847/1538-4357/ac3b4c}, \href {https://ui.adsabs.harvard.edu/abs/2022ApJ...926...31R} {926, 31}

\bibitem[\protect\citeauthoryear{{Reddy} et~al.,}{{Reddy} et~al.}{2023a}]{Reddy2023a}
{Reddy} N.~A.,  et~al., 2023a, \mn@doi [\apj] {10.3847/1538-4357/acd0b1}, \href {https://ui.adsabs.harvard.edu/abs/2023ApJ...951...56R} {951, 56}

\bibitem[\protect\citeauthoryear{{Reddy}, {Topping}, {Sanders}, {Shapley}  \& {Brammer}}{{Reddy} et~al.}{2023b}]{Reddy2023b}
{Reddy} N.~A.,  {Topping} M.~W.,  {Sanders} R.~L.,  {Shapley} A.~E.,   {Brammer} G.,  2023b, \mn@doi [\apj] {10.3847/1538-4357/acd754}, \href {https://ui.adsabs.harvard.edu/abs/2023ApJ...952..167R} {952, 167}

\bibitem[\protect\citeauthoryear{{Roberts-Borsani} et~al.,}{{Roberts-Borsani} et~al.}{2024}]{RobertsBorsani2024}
{Roberts-Borsani} G.,  et~al., 2024, \mn@doi [arXiv e-prints] {10.48550/arXiv.2403.07103}, \href {https://ui.adsabs.harvard.edu/abs/2024arXiv240307103R} {p. arXiv:2403.07103}

\bibitem[\protect\citeauthoryear{{Sanders} et~al.,}{{Sanders} et~al.}{2016a}]{Sanders2016}
{Sanders} R.~L.,  et~al., 2016a, \mn@doi [\apj] {10.3847/0004-637X/816/1/23}, \href {https://ui.adsabs.harvard.edu/abs/2016ApJ...816...23S} {816, 23}

\bibitem[\protect\citeauthoryear{{Sanders} et~al.,}{{Sanders} et~al.}{2016b}]{Sanders2016a}
{Sanders} R.~L.,  et~al., 2016b, \mn@doi [\apj] {10.3847/0004-637X/816/1/23}, \href {https://ui.adsabs.harvard.edu/abs/2016ApJ...816...23S} {816, 23}

\bibitem[\protect\citeauthoryear{{Sanders} et~al.,}{{Sanders} et~al.}{2021}]{Sanders2021}
{Sanders} R.~L.,  et~al., 2021, \mn@doi [\apj] {10.3847/1538-4357/abf4c1}, \href {https://ui.adsabs.harvard.edu/abs/2021ApJ...914...19S} {914, 19}

\bibitem[\protect\citeauthoryear{{Sanders} et~al.,}{{Sanders} et~al.}{2024a}]{Sanders2024}
{Sanders} R.~L.,  et~al., 2024a, \mn@doi [arXiv e-prints] {10.48550/arXiv.2408.05273}, \href {https://ui.adsabs.harvard.edu/abs/2024arXiv240805273S} {p. arXiv:2408.05273}

\bibitem[\protect\citeauthoryear{{Sanders}, {Shapley}, {Topping}, {Reddy}  \& {Brammer}}{{Sanders} et~al.}{2024b}]{Sanders2024b}
{Sanders} R.~L.,  {Shapley} A.~E.,  {Topping} M.~W.,  {Reddy} N.~A.,   {Brammer} G.~B.,  2024b, \mn@doi [\apj] {10.3847/1538-4357/ad15fc}, \href {https://ui.adsabs.harvard.edu/abs/2024ApJ...962...24S} {962, 24}

\bibitem[\protect\citeauthoryear{{Senchyna}, {Plat}, {Stark}, {Rudie}, {Berg}, {Charlot}, {James}  \& {Mingozzi}}{{Senchyna} et~al.}{2024}]{Senchyna2024}
{Senchyna} P.,  {Plat} A.,  {Stark} D.~P.,  {Rudie} G.~C.,  {Berg} D.,  {Charlot} S.,  {James} B.~L.,   {Mingozzi} M.,  2024, \mn@doi [\apj] {10.3847/1538-4357/ad235e}, \href {https://ui.adsabs.harvard.edu/abs/2024ApJ...966...92S} {966, 92}

\bibitem[\protect\citeauthoryear{{Shapley} et~al.,}{{Shapley} et~al.}{2017}]{Shapley2017}
{Shapley} A.~E.,  et~al., 2017, \mn@doi [\apjl] {10.3847/2041-8213/aa8815}, \href {https://ui.adsabs.harvard.edu/abs/2017ApJ...846L..30S} {846, L30}

\bibitem[\protect\citeauthoryear{{Shapley}, {Reddy}, {Sanders}, {Topping}  \& {Brammer}}{{Shapley} et~al.}{2023}]{Shapley2023}
{Shapley} A.~E.,  {Reddy} N.~A.,  {Sanders} R.~L.,  {Topping} M.~W.,   {Brammer} G.~B.,  2023, \mn@doi [\apjl] {10.3847/2041-8213/acd939}, \href {https://ui.adsabs.harvard.edu/abs/2023ApJ...950L...1S} {950, L1}

\bibitem[\protect\citeauthoryear{{Shapley} et~al.,}{{Shapley} et~al.}{2024}]{Shapley2024}
{Shapley} A.~E.,  et~al., 2024, \mn@doi [arXiv e-prints] {10.48550/arXiv.2407.00157}, \href {https://ui.adsabs.harvard.edu/abs/2024arXiv240700157S} {p. arXiv:2407.00157}

\bibitem[\protect\citeauthoryear{{Shimakawa} et~al.,}{{Shimakawa} et~al.}{2015}]{Shimakawa2015}
{Shimakawa} R.,  et~al., 2015, \mn@doi [\mnras] {10.1093/mnras/stv915}, \href {https://ui.adsabs.harvard.edu/abs/2015MNRAS.451.1284S} {451, 1284}

\bibitem[\protect\citeauthoryear{{Shirazi}, {Brinchmann}  \& {Rahmati}}{{Shirazi} et~al.}{2014}]{Shirazi2014}
{Shirazi} M.,  {Brinchmann} J.,   {Rahmati} A.,  2014, \mn@doi [\apj] {10.1088/0004-637X/787/2/120}, \href {https://ui.adsabs.harvard.edu/abs/2014ApJ...787..120S} {787, 120}

\bibitem[\protect\citeauthoryear{{Speagle}, {Steinhardt}, {Capak}  \& {Silverman}}{{Speagle} et~al.}{2014}]{Speagle2014}
{Speagle} J.~S.,  {Steinhardt} C.~L.,  {Capak} P.~L.,   {Silverman} J.~D.,  2014, \mn@doi [\apjs] {10.1088/0067-0049/214/2/15}, \href {https://ui.adsabs.harvard.edu/abs/2014ApJS..214...15S} {214, 15}

\bibitem[\protect\citeauthoryear{{Stanway} \& {Eldridge}}{{Stanway} \& {Eldridge}}{2018}]{Stanway2018}
{Stanway} E.~R.,  {Eldridge} J.~J.,  2018, \mn@doi [\mnras] {10.1093/mnras/sty1353}, \href {https://ui.adsabs.harvard.edu/abs/2018MNRAS.479...75S} {479, 75}

\bibitem[\protect\citeauthoryear{{Steidel} et~al.,}{{Steidel} et~al.}{2014}]{Steidel2014}
{Steidel} C.~C.,  et~al., 2014, \mn@doi [\apj] {10.1088/0004-637X/795/2/165}, \href {https://ui.adsabs.harvard.edu/abs/2014ApJ...795..165S} {795, 165}

\bibitem[\protect\citeauthoryear{{Steidel}, {Strom}, {Pettini}, {Rudie}, {Reddy}  \& {Trainor}}{{Steidel} et~al.}{2016}]{Steidel2016}
{Steidel} C.~C.,  {Strom} A.~L.,  {Pettini} M.,  {Rudie} G.~C.,  {Reddy} N.~A.,   {Trainor} R.~F.,  2016, \mn@doi [\apj] {10.3847/0004-637X/826/2/159}, \href {https://ui.adsabs.harvard.edu/abs/2016ApJ...826..159S} {826, 159}

\bibitem[\protect\citeauthoryear{{Tacchella} et~al.,}{{Tacchella} et~al.}{2022}]{Tacchella2021}
{Tacchella} S.,  et~al., 2022, \mn@doi [\apj] {10.3847/1538-4357/ac4cad}, \href {https://ui.adsabs.harvard.edu/abs/2022ApJ...927..170T} {927, 170}

\bibitem[\protect\citeauthoryear{{Tang}, {Stark}, {Chevallard}  \& {Charlot}}{{Tang} et~al.}{2019}]{Tang2019}
{Tang} M.,  {Stark} D.~P.,  {Chevallard} J.,   {Charlot} S.,  2019, \mn@doi [\mnras] {10.1093/mnras/stz2236}, \href {https://ui.adsabs.harvard.edu/abs/2019MNRAS.489.2572T} {489, 2572}

\bibitem[\protect\citeauthoryear{{Tang} et~al.,}{{Tang} et~al.}{2023}]{Tang2023}
{Tang} M.,  et~al., 2023, \mn@doi [\mnras] {10.1093/mnras/stad2763}, \href {https://ui.adsabs.harvard.edu/abs/2023MNRAS.526.1657T} {526, 1657}

\bibitem[\protect\citeauthoryear{{Topping} et~al.,}{{Topping} et~al.}{2022a}]{Topping2022a}
{Topping} M.~W.,  et~al., 2022a, \mn@doi [\mnras] {10.1093/mnras/stac2291}, \href {https://ui.adsabs.harvard.edu/abs/2022MNRAS.516..975T} {516, 975}

\bibitem[\protect\citeauthoryear{{Topping}, {Stark}, {Endsley}, {Plat}, {Whitler}, {Chen}  \& {Charlot}}{{Topping} et~al.}{2022b}]{Topping2022}
{Topping} M.~W.,  {Stark} D.~P.,  {Endsley} R.,  {Plat} A.,  {Whitler} L.,  {Chen} Z.,   {Charlot} S.,  2022b, \mn@doi [\apj] {10.3847/1538-4357/aca522}, \href {https://ui.adsabs.harvard.edu/abs/2022ApJ...941..153T} {941, 153}

\bibitem[\protect\citeauthoryear{{Topping} et~al.,}{{Topping} et~al.}{2024a}]{Topping2024c}
{Topping} M.~W.,  et~al., 2024a, \mn@doi [arXiv e-prints] {10.48550/arXiv.2407.19009}, \href {https://ui.adsabs.harvard.edu/abs/2024arXiv240719009T} {p. arXiv:2407.19009}

\bibitem[\protect\citeauthoryear{{Topping} et~al.,}{{Topping} et~al.}{2024b}]{Topping2024a}
{Topping} M.~W.,  et~al., 2024b, \mn@doi [\mnras] {10.1093/mnras/stae682}, \href {https://ui.adsabs.harvard.edu/abs/2024MNRAS.529.3301T} {529, 3301}

\bibitem[\protect\citeauthoryear{{Tremonti} et~al.,}{{Tremonti} et~al.}{2004}]{Tremonti2004}
{Tremonti} C.~A.,  et~al., 2004, \mn@doi [\apj] {10.1086/423264}, \href {https://ui.adsabs.harvard.edu/abs/2004ApJ...613..898T} {613, 898}

\bibitem[\protect\citeauthoryear{{Whitler}, {Stark}, {Endsley}, {Leja}, {Charlot}  \& {Chevallard}}{{Whitler} et~al.}{2023}]{Whitler2023}
{Whitler} L.,  {Stark} D.~P.,  {Endsley} R.,  {Leja} J.,  {Charlot} S.,   {Chevallard} J.,  2023, \mn@doi [\mnras] {10.1093/mnras/stad004}, \href {https://ui.adsabs.harvard.edu/abs/2023MNRAS.519.5859W} {519, 5859}

\bibitem[\protect\citeauthoryear{{Witstok}, {Smit}, {Maiolino}, {Curti}, {Laporte}, {Massey}, {Richard}  \& {Swinbank}}{{Witstok} et~al.}{2021}]{Witstok2021}
{Witstok} J.,  {Smit} R.,  {Maiolino} R.,  {Curti} M.,  {Laporte} N.,  {Massey} R.,  {Richard} J.,   {Swinbank} M.,  2021, \mn@doi [\mnras] {10.1093/mnras/stab2591}, \href {https://ui.adsabs.harvard.edu/abs/2021MNRAS.508.1686W} {508, 1686}

\bibitem[\protect\citeauthoryear{{Yanagisawa} et~al.,}{{Yanagisawa} et~al.}{2024}]{Yanagisawa2024}
{Yanagisawa} H.,  et~al., 2024, \mn@doi [\apj] {10.3847/1538-4357/ad72ec}, \href {https://ui.adsabs.harvard.edu/abs/2024ApJ...974..266Y} {974, 266}

\bibitem[\protect\citeauthoryear{{de Graaff} et~al.,}{{de Graaff} et~al.}{2024}]{deGraaf2023}
{de Graaff} A.,  et~al., 2024, \mn@doi [\aap] {10.1051/0004-6361/202347755}, \href {https://ui.adsabs.harvard.edu/abs/2024A&A...684A..87D} {684, A87}

\makeatother
\end{thebibliography}
\appendix
\section{Table of Electron Densities}
Table~\ref{tab:densities} presents the electron densities for our sample. The selection of electron density sample is described in Section~\ref{sec:densities}, and the methodology used to derive electron densities based on the [SII] and CIII] doublet ratios is described in Section~\ref{sec:sample}.

\begin{table*}
\caption{Objects in our electron density sample (see Section~\ref{sec:densities}). This table presents object IDs, coordinates, and spectroscopic redshifts, in addition to the [SII] and CIII] line ratios (if detected) with their associated electron densities.}
\begin{adjustbox}{width=1.0\linewidth,center=1\linewidth}
\renewcommand{\arraystretch}{1.2}
\hskip-0.45cm\begin{tabular}{lccccccc}
\toprule
ID & R.A. & Decl. & $z_{\rm spec}$ & $\rm [SII](6717/6730)$ & $n_{\rm e}(\rm [SII])$ & $\rm CIII](1907/1909)$ & $n_{\rm e}(\rm CIII])$ \\
 & (J2000)  &  (J2000) &  & & $\rm [cm^{-3}]$ &  & $\rm [cm^{-3}]$ \\
\midrule
GOODSN-19149 & 189.13610 & 62.23648 & $1.3833$ & $1.21^{+0.10}_{-0.10}$ & $276^{+192}_{-137}$ & $-$ & $-$ \\
COSMOS-9316 & 150.17098 & 2.27436 & $1.4159$ & $1.17^{+0.07}_{-0.08}$ & $334^{+141}_{-99}$ & $-$ & $-$ \\
COSMOS-7627 & 150.16509 & 2.25666 & $1.4968$ & $1.23^{+0.06}_{-0.06}$ & $247^{+98}_{-76}$ & $-$ & $-$ \\
GOODSN-929174 & 189.13467 & 62.27702 & $1.5218$ & $1.45^{+0.06}_{-0.06}$ & $37^{+42}_{-28}$ & $-$ & $-$ \\
COSMOS-4622 & 150.15468 & 2.22459 & $1.6400$ & $0.97^{+0.06}_{-0.06}$ & $794^{+195}_{-177}$ & $-$ & $-$ \\
GOODSN-30274 & 189.18297 & 62.29196 & $1.7997$ & $1.21^{+0.39}_{-0.39}$ & $518^{+1776}_{-394}$ & $-$ & $-$ \\
COSMOS-4205 & 150.13992 & 2.22020 & $1.8368$ & $1.29^{+0.05}_{-0.04}$ & $164^{+57}_{-55}$ & $-$ & $-$ \\
COSMOS-5870 & 150.18118 & 2.23752 & $1.8429$ & $1.17^{+0.18}_{-0.14}$ & $343^{+288}_{-238}$ & $-$ & $-$ \\
COSMOS-5826 & 150.17966 & 2.23712 & $1.9256$ & $1.37^{+0.16}_{-0.17}$ & $144^{+166}_{-99}$ & $-$ & $-$ \\
COSMOS-8467 & 150.13813 & 2.26574 & $1.9609$ & $1.46^{+0.17}_{-0.23}$ & $174^{+198}_{-113}$ & $-$ & $-$ \\
GOODSN-929125 & 189.15135 & 62.27704 & $1.9907$ & $1.20^{+0.27}_{-0.25}$ & $431^{+691}_{-312}$ & $-$ & $-$ \\
GOODSN-25004 & 189.19237 & 62.26435 & $2.0487$ & $1.16^{+0.03}_{-0.04}$ & $358^{+70}_{-46}$ & $-$ & $-$ \\
COSMOS-446062 & 150.14007 & 2.26315 & $2.0638$ & $1.27^{+0.13}_{-0.14}$ & $191^{+216}_{-138}$ & $-$ & $-$ \\
COSMOS-4029 & 150.15277 & 2.21873 & $2.0765$ & $1.30^{+0.06}_{-0.06}$ & $157^{+78}_{-69}$ & $-$ & $-$ \\
COSMOS-5161 & 150.13804 & 2.22993 & $2.1015$ & $0.79^{+0.23}_{-0.19}$ & $1627^{+2030}_{-836}$ & $-$ & $-$ \\
COSMOS-4429 & 150.13616 & 2.22248 & $2.1023$ & $1.20^{+0.16}_{-0.15}$ & $295^{+268}_{-180}$ & $-$ & $-$ \\
COSMOS-5814 & 150.16908 & 2.23839 & $2.1269$ & $1.15^{+0.07}_{-0.06}$ & $377^{+117}_{-114}$ & $-$ & $-$ \\
COSMOS-7883 & 150.13582 & 2.25994 & $2.1532$ & $1.26^{+0.08}_{-0.07}$ & $211^{+104}_{-91}$ & $-$ & $-$ \\
COSMOS-5283 & 150.18665 & 2.23196 & $2.1742$ & $1.23^{+0.03}_{-0.02}$ & $249^{+30}_{-38}$ & $-$ & $-$ \\
COSMOS-4156 & 150.17931 & 2.21977 & $2.1897$ & $1.15^{+0.05}_{-0.06}$ & $371^{+107}_{-89}$ & $-$ & $-$ \\
GOODSN-30053 & 189.19754 & 62.29132 & $2.2454$ & $1.20^{+0.03}_{-0.03}$ & $299^{+44}_{-50}$ & $-$ & $-$ \\
GOODSN-27876 & 189.12378 & 62.27925 & $2.2709$ & $1.04^{+0.05}_{-0.04}$ & $600^{+106}_{-112}$ & $-$ & $-$ \\
COSMOS-5571 & 150.19130 & 2.23591 & $2.2784$ & $1.29^{+0.05}_{-0.04}$ & $164^{+51}_{-61}$ & $-$ & $-$ \\
GOODSN-30811 & 189.17584 & 62.29549 & $2.3067$ & $1.06^{+0.08}_{-0.09}$ & $565^{+233}_{-161}$ & $-$ & $-$ \\
COSMOS-3324 & 150.14840 & 2.21314 & $2.3077$ & $1.33^{+0.10}_{-0.08}$ & $128^{+101}_{-87}$ & $-$ & $-$ \\
GOODSN-21522 & 189.20375 & 62.24755 & $2.3634$ & $1.31^{+0.05}_{-0.05}$ & $142^{+63}_{-57}$ & $-$ & $-$ \\
COSMOS-5901 & 150.18939 & 2.23814 & $2.3966$ & $1.35^{+0.05}_{-0.05}$ & $106^{+63}_{-52}$ & $-$ & $-$ \\
GOODSN-22235 & 189.13922 & 62.25133 & $2.4298$ & $1.16^{+0.03}_{-0.02}$ & $354^{+41}_{-49}$ & $-$ & $-$ \\
GOODSN-30564 & 189.20408 & 62.29434 & $2.4828$ & $1.05^{+0.03}_{-0.02}$ & $586^{+54}_{-60}$ & $-$ & $-$ \\
GOODSN-26798 & 189.15458 & 62.27462 & $2.4831$ & $1.09^{+0.03}_{-0.02}$ & $485^{+54}_{-54}$ & $-$ & $-$ \\
GOODSN-931951 & 189.19812 & 62.28700 & $2.4834$ & $1.25^{+0.14}_{-0.14}$ & $217^{+239}_{-148}$ & $-$ & $-$ \\
GOODSN-917938 & 189.13919 & 62.24336 & $2.9219$ & $1.06^{+0.14}_{-0.15}$ & $568^{+444}_{-254}$ & $-$ & $-$ \\
COSMOS-8813 & 150.16442 & 2.26922 & $2.9262$ & $1.36^{+0.08}_{-0.07}$ & $93^{+71}_{-73}$ & $-$ & $-$ \\
GOODSN-919341 & 189.23045 & 62.24698 & $2.9586$ & $0.92^{+0.33}_{-0.27}$ & $957^{+1687}_{-646}$ & $-$ & $-$ \\
GOODSN-19848 & 189.16046 & 62.23943 & $2.9918$ & $0.92^{+0.05}_{-0.05}$ & $994^{+215}_{-187}$ & $-$ & $-$ \\
GOODSN-914864 & 189.17240 & 62.23499 & $2.9919$ & $0.90^{+0.19}_{-0.19}$ & $1048^{+1333}_{-572}$ & $-$ & $-$ \\
GOODSN-22384 & 189.16149 & 62.25202 & $2.9935$ & $1.27^{+0.04}_{-0.04}$ & $191^{+56}_{-49}$ & $-$ & $-$ \\
COSMOS-4113 & 150.13793 & 2.22025 & $3.0849$ & $1.15^{+0.06}_{-0.06}$ & $382^{+121}_{-94}$ & $-$ & $-$ \\
GOODSN-21033 & 189.21160 & 62.24571 & $3.1120$ & $1.39^{+0.08}_{-0.07}$ & $55^{+78}_{-49}$ & $-$ & $-$ \\
COSMOS-4740 & 150.15878 & 2.22606 & $3.1555$ & $1.02^{+0.03}_{-0.03}$ & $653^{+69}_{-86}$ & $-$ & $-$ \\
\bottomrule
\end{tabular}
\end{adjustbox}
\label{tab:densities}
\end{table*}
\begin{table*}
\caption{Continued}
\begin{adjustbox}{width=1.0\linewidth,center=1\linewidth}
\renewcommand{\arraystretch}{1.2}
\hskip-0.45cm\begin{tabular}{lccccccc}
\toprule
ID & R.A. & Decl. & $z_{\rm spec}$ & $\rm [SII](6717/6730)$ & $n_{\rm e}(\rm [SII])$ & $\rm CIII](1907/1909)$ & $n_{\rm e}(\rm CIII])$ \\
 &  &  &  & & $\rm [cm^{-3}]$ &  & $\rm [cm^{-3}]$ \\
\midrule

COSMOS-8697 & 150.16569 & 2.26818 & $3.2068$ & $1.35^{+0.31}_{-0.24}$ & $234^{+401}_{-146}$ & $-$ & $-$ \\
GOODSN-28209 & 189.18842 & 62.28118 & $3.2325$ & $0.89^{+0.06}_{-0.07}$ & $1091^{+331}_{-232}$ & $-$ & $-$ \\
COSMOS-8363 & 150.13641 & 2.26468 & $3.2475$ & $1.19^{+0.06}_{-0.05}$ & $303^{+77}_{-83}$ & $-$ & $-$ \\
GOODSN-22932 & 189.21601 & 62.25436 & $3.3305$ & $1.26^{+0.13}_{-0.13}$ & $214^{+196}_{-129}$ & $-$ & $-$ \\
GOODSN-23927 & 189.24105 & 62.25951 & $3.3643$ & $0.95^{+0.13}_{-0.12}$ & $871^{+500}_{-318}$ & $-$ & $-$ \\
GOODSN-21726 & 189.13937 & 62.24906 & $3.4090$ & $1.25^{+0.32}_{-0.45}$ & $395^{+1612}_{-286}$ & $-$ & $-$ \\
COSMOS-6124 & 150.18166 & 2.23996 & $3.4589$ & $0.98^{+0.06}_{-0.06}$ & $781^{+218}_{-172}$ & $-$ & $-$ \\
GOODSN-17940 & 189.14785 & 62.23057 & $4.4115$ & $0.97^{+0.04}_{-0.03}$ & $808^{+104}_{-108}$ & $-$ & $-$ \\
GOODSN-917107 & 189.18228 & 62.24138 & $4.7729$ & $-$ & $-$ & $0.38^{+0.09}_{-0.10}$ & $103500^{+49700}_{-26800}$ \\
GOODSN-100067 & 189.23479 & 62.25748 & $5.1875$ & $1.42^{+0.34}_{-0.35}$ & $237^{+823}_{-163}$ & $1.01^{+0.12}_{-0.11}$ & $17400^{+6400}_{-5400}$ \\
COSMOS-443467 & 150.14566 & 2.25855 & $5.5039$ & $-$ & $-$ & $1.33^{+0.30}_{-0.25}$ & $4700^{+11400}_{-4500}$ \\
COSMOS-440430 & 150.14807 & 2.25324 & $5.5207$ & $0.88^{+0.26}_{-0.34}$ & $1125^{+2244}_{-538}$ & $-$ & $-$ \\
COSMOS-3661 & 150.15863 & 2.21509 & $5.8227$ & $1.04^{+0.27}_{-0.33}$ & $594^{+1424}_{-381}$ & $0.79^{+0.18}_{-0.18}$ & $31900^{+19600}_{-12400}$ \\
GOODSN-100163 & 189.13811 & 62.27444 & $6.7479$ & $-$ & $-$ & $1.44^{+0.28}_{-0.30}$ & $6400^{+8100}_{-3900}$ \\
GOODSN-100167 & 189.20259 & 62.27553 & $6.9063$ & $-$ & $-$ & $0.97^{+0.28}_{-0.18}$ & $19400^{+12800}_{-13600}$ \\
GOODSN-100026 & 189.22513 & 62.28629 & $7.2043$ & $-$ & $-$ & $1.40^{+0.34}_{-0.39}$ & $8100^{+15900}_{-6200}$ \\
GOODSN-511948 & 189.23379 & 62.26397 & $10.3861$ & $-$ & $-$ & $1.04^{+0.26}_{-0.30}$ & $16200^{+21300}_{-10700}$ \\
\bottomrule
\end{tabular}
\end{adjustbox}
\label{tab:densitiescont}
\end{table*}

\end{document}